# A Global and Local Structure-Based Method for Predicting Binary Protein-Protein Interaction Partners: Proof of Principle and Feasibility

**Vicente M. Reyes, Ph.D.***
**E-mail: vmrsbi.RIT.biology@gmail.com**

*work done at:

Dept. of Pharmacology, School of Medicine,
University of California, San Diego
9500 Gilman Drive, La Jolla, CA 92093-0636
&
Dept. of Biological Sciences, School of Life Sciences
Rochester Institute of Technology
One Lomb Memorial Drive, Rochester, NY 14623

**Abstract**. We report a novel 3D structure-based method of predicting protein-protein (PP) interaction (PPI) partners. The method involves screening for pairs of tetrahedra representing interacting amino acids at the interface of the PP complex, with one tetrahedron on each protomer*. H-bonds and VDW interactions at the interface of the two interacting protein molecules in the complex are first determined, and then interacting tetrahedral motifs - one from each protomer – representing backbone or side chain centroids of the interacting amino acids, are built from them. The method requires that the protein protomers be transformed first into double-centroid reduced representation (DCRR; Reyes, V.M. & Sheth, V.N., 2011; Reyes, V.M., 2015a). The method is applied to a collection of 801 functionally unannotated protein structures in the Protein Data Bank (PDB), which were screened for pairs of tetrahedral motifs characteristic of nine binary complexes, namely: (1.) RAP-Gmppnp - c-RAF1 Ras-binding domain; (2.) RHOA - protein kinase PKN/PRK1 effector domain; (3.) RAC - RHOGD1; (4.) RAC - P67PHOX; (5.) kinase-associated phosphatase (KAP) - phospho-CDK2; (6.) Ig Fc – protein A fragment B; (7.) Ig light chain dimers; (8.) beta-catenin - HTCF-4; and (9.) IL-2 homodimers. Our search procedure found 33, 297, 62, 63, 120, 0, 108, 16 and 504 putative complexes, respectively. After considering the degree of interfacial overlap between the two protomers in the binary complex, these numbers were significantly trimmed down to 4, 2, 1, 8, 3, 0, 1, 1, and 1, respectively. Negative and positive control experiments indicate that the screening process has better than acceptable specificity and sensitivity. The results were further validated by applying the "Cutting Plane" and 'Tangent Sphere" methods (Reyes, V.M., 2015b) for the quantitative determination of interface burial, which is indicative of inter-protomer overlap (IPO) in the binary complex; which we, in turn, assume to be a conserved property of any specific family of binary complexes. One advantage of our method is its simplicity, speed and scalability, as well as its protein docking nature -- i.e., once the partner interface 3D SMs are identified, they can be used to computationally dock the two protomers together to form the putative binary complex -- a property that we shall describe briefly in this work.

---

***We shall use the term "protomer" to refer to each of the two monomer molecules making up the binary protein complex.**

**Abbreviations**: DCRR, double-centroid residue representation; X(b), the backbone centroid of amino acid X; Z(s), the side-chain centroid of amino acid Z; PDB, Protein Data Bank; HB, hydrogen-bond(ing); VDW, van der Waals (interactions); PLI, protein-ligand interactions; PPI, protein-protein interactions; BS, binding site; LBS, ligand binding site; CP(M), cutting plane (method); CPi, cutting plane index; TS(M), tangent sphere (method); TSi, tangent sphere index; FUA, functionally unannotated (with unknown function); NNA, nearest-neigbor analysis; 3D SM, three-dimensional search motif; 3D ISMTP, 3D interface search motif tetrahedral pair; BC, binary complex; IPO, inter-protomer overlap (at interface of complex); RMSD, root-mean-square deviation.

## 1 Introduction:

### 1.1 Motivation and Background.

The current proliferation of genomic sequencing and high-throughput structural genomics studies around the world (Burley, S.K. 2000; Heinemann, U. 2000; Terwilliger, T.C., 2000; Norrvell, J.C., & Machalek, A.Z., 2000) have necessitated a great need for computational methods of protein function assignment, especially those based on protein three-dimensional (3D) structure (Bentley et al., 2004; Fernandez-Ballester et al., 2006; Murphy et al, 2004; Baxevanis, 2003; Miller et al,, 2003; Jung, J.W. & Lee, W. 2004; Yakunin, A.F., et al., 2004). Two general approaches to function assignment are the prediction of protein-ligand and protein-protein interactions (PLI and PPI, respectively; Burgoyne et al., 2006; Deng et al., 2003). This work deals with the prediction of PPI partners. The method was originally applied to, and is an extension of, the prediction of PLI, where tetrahedral 3D motifs in proteins representing amino acids in the ligand binding sites (LBS) are screened for (Reyes, V.M. & Sheth, V.N., 2011; Reyes, V.M., 2015a). In the case of PPI, there is not one but two tetrahedral 3D search motifs (3D SM); one comes from the first protomer and the other from the second, and both are at their interface in the complex. A schematic diagram of interacting amino acids at the interface of a protein binary complex (BC) is shown in Figure 1. This study is limited to binary protein complexes; higher-order protein complexes will be addressed in a future work by us.

Currently, there are several available methods of computational PPI prediction (for excellent reviews, see Shi et al., 2005; Russell et al., 2004; Obenauer et al., 2004; Valencia et al., 2002). Some rely on statistical machine learning techniques such as support vector machines (SVM; Lo et al., 2005) or neural networks (Fariselli et al., 2002) and thus are purely stochastic or probabilistic (Han et al., 2003; Lu et al., 2003); others rely alone on sequence data, and are thus insufficient. Still, others specify only that a surface patch (Jones & Thornton, 1997) on the query protein is a probable protein interaction site, without specifying the identity of the interaction partner, thus precluding complete functional annotation. Other computational methods of PPI prediction are based on geometric description and/or electrostatic characterization of protein surfaces (Kundrotas & Alexov, 2006), followed by screening for matches between sites on these surfaces. Screening is usually performed by constructing grids of arbitrary size, and matches searched by pairwise and iterative comparison of such grids. Most of these methods are computationally uneconomical and not amenable to large-scale implementation. Some methods rely on computer-intensive molecular simulations techniques, and are therefore not amenable to large-scale, high-throughout implementation, as is required for proteomics work. Our present method overcomes some of these above obstacles.

### 1.2 Overview of the Method.

Our screening method is essentially based on the interactions at the interface of the two protomers when docked together as a complex (Li et al., 2006). The overall procedure is shown schematically in Figure 2. We limit our work to BCs at this time, postponing addressing tertiary, quarternary and other higher-order complexes in future work, enough to say at this point that treatment of these higher-order complexes will involve considering the protomers two at a time, i.e., pairwise.

We focus our attention on the PP interface because it is where the complex is stabilized and it is also well established that the interface possesses certain special properties which are conserved (Bernauer et al., 2007; Block et al., 2006; Winter et al., 2006; Dong et al., 2004; Mintseris et al., 2003). The procedure starts with the determination of the HB and VDW interactions at the interface of an experimentally solved training structure for the complex under study. This is accomplished by running a nearest neighbor analysis (NNA) program we have written, followed by another program that selects HB and VDW interactions from the neighbors obtained (Reyes, V.M. & Sheth, V.N., 2011; Reyes, V.M., 2015a; Bondi, 1964; Engh & Huber, 1991). Then a tetrahedral 3D SM from each protomer is constructed from these interfacial interactions, giving rise to an interacting ("docked") pair of 3D SM's (Figure 3). The "application set" - the set of proteins to be functionally annotated -- is then screened for each 3D SM, and those testing positive for either motif are deemed candidate PPI partners.

The results are validated by determining the depth of burial of the 3D SM within each protomer, as well as the degree of overlap between the two candidate protomer partners – a quantity we term as inter-protomer overlap (IPO) - when docked together using their 3D SMs. Additionally, our procedure may also performed in the undocked mode, as described in the Methods section. Comparison of the IPO from the candidate application structures with those determined from their respective training structure(s) allows us to weed out candidate structures that are less likely to be true positives than others. The procedures we used for determining the 3D SM burial and IPO are the "cutting plane" (CP) and "tangent sphere" (TS) methods, two novel complementary procedures presented in an accompanying paper (see Figure 1A of Reyes, V.M., 2015b). The CP and TS methods complement each other, as the CP method is directly proportional to the 3D SM burial depth, while the TS method is inversely proportional to it. Our method described here is, to our knowledge, one of the first analytical methods devised for PPI partner prediction. Furthermore, it allows specific function prediction, as it is specific for the PPI partners under consideration.

## 2 Datasets and Methods.

### 2.1 Datasets.

The training structures were obtained from the PDB (Berman, H.M., & Westbrook, J.D. 2004) by keyword search for "protein-protein complex." Of the hundreds of structures outputted, those which were binary (made up of exactly two protein chains) and of human origin were selected. The set was further narrowed down by selecting only those which did not have mutations, have reasonable resolutions, and did not have too many chains in the structure. The 9 training structures (1 for each complex) are shown in Table 1; they are: (1.) 1c1y: RAP-Gmppnp/c-RAF1 Ras-binding protein; (2.) 1cxz: RHOA/Protein Kinase (PKN/PRK1) effector domain; (3.) 1ds6: RAC/RHOGD1; (4.) 1e96: RAC/P67PHOX; (5.) 1fq1: kinase-associated phosphatase (KAP)/phospho-CDK2; (6.) 1fc2: immunoglobulin Fc fragment/protein A fragment B; (7.) 1mco: immunoglobulin light chain dimer (Bence-Jones protein); (8.) 1jdh: beta-catenin/HTCF-4; and (9.) 3ink: interleukin-2 homodimer. They are designated as complexes A, B, C, D, H, I, P, Q and Z, respectively.

#### 2.1.1 Positive and Negative Control Tests.

The negative control structures used in determining the specificity of the screening procedure for all nine binary complexes are enumerated and described in Table 3 of Reyes, V.M., 2015c. Lack of positive control structures in the PDB to determine the sensitivity of the method was circumvented by creating artificial binding sites (BSs) corresponding to those determined in the nine training structure BCs by replacing relevant amino acids in some of the negative control structures. Negative control experiments showed that the method has ample specificity, while positive control experiments demonstrated that it has ample sensitivity.

#### 2.1.2 The Application Set: 801 Functionally Unannotated Protein Structures in the PDB

The 801 application structures are shown in Table 2 (parts 1 and 2) of Reyes, V.M., 2015c. This "application set" was obtained from the PDB by keyword search for "unknown function" or a similar phrase, then verified by "grepping" (i.e., using the UNIX "grep" command) the resulting PDB files'

"HEADER" record and ascertaining that the descriptor "UNKNOWN FUNCTION" is indeed present. They were further verified by running an SQL script that searches for structures without any GO term annotations nor E.C. assignment. The initial set was narrowed down by excluding those which were not solved by x-ray crystallography. The final result is a set composed of 801 x-ray crystallographically solved structures which do not currently have functional annotations.

The species distribution of the 801 FUA application protein structures is shown and described in Table 4 of Reyes, V.M., 2015c. They come from 104 known species (which include bacteria, archaea, protozoans, and some higher organisms including humans), an uncultured bacterium (unknown species), and one is a synthetic protein.

## 2.2 Methods.

The PP complexes we selected as training structures have at least one structure experimentally solved and whose coordinates are deposited in the PDB. In this work, we focus on binary complexes; higher-order complexes have exponentially greater degree of screening complexity and will be addressed in our future work in this area. Steps 1 and 2 below deal with the training structures; succeeding steps deal with the "application structures" – proteins whose 3D structures are known (experimentally solved) but whose function(s) are unknown – that we wish to assign function to. All calculations are done using Fortran 77 or 90 programs and/or UNIX utilities and scripts. The overall procedure is shown schematically in Figure 2.

Our procedure for predicting PPI partners is composed of five steps, namely, (1.) NNA; (2) determination of protomer interface HB and VDW interactions; (3.) determination of the interface tetrahedral search motif pair; (4.) screening the application set for the tetrahedral search motif pair; and (5.) validation of the putative protomer partners by determining the IPO in the complex using the cutting plane and tangent sphere methods. Each step is described below.

### 2.2.1 Overview of the Elimination Process.

The multi-step elimination process we employed is shown in Figure 4. Both local and global structure information are utilized in the process to search for the best candidate positive structures. Set A, the outermost pink circle, represents the starting application set composed of 801 PDB structures without functional annotation. These are then screened for the particular 3D SM in question, and those that test positive form a subset of A, which we call set B (blue area), which are structures where at least one 3D SM was detected. Set B structures are then subjected to the "Cutting Plane" and "Tangent sphere" Methods (Reyes, V.M., 2015b) and their CP and TS indices (CPi and TSi, respectively) compared respectively to those of the training structures used to create the 3D SMs. Those whose CPi or TSi are close enough (typically within 8-10 %) of those of the training structures are selected to form a subset of B which we call set C (tan area). Structures in set C are further analyzed to determine whether their indices are both respectively similar to those of (a) specific training structures. Those which satisfy this criterion form a subset of C, which we call set D (green area). Finally, those whose IPO between protomers are close to those in the training structure form set E (purple circle in center). We now describe each of these steps in more detail below.

### 2.2.1 Step 1: Determination of Protomer Interface HB and VDW Interactions.

This step deals with the training structures in all-atom representation (AAR; all atoms except hydrogens have coordinates). This step is the same as the determination of the LBS consensus motif in our previous work (Reyes, V.M., 2007a, c), except that the "ligand" of protomer1 is protomer2, and that of protomer #2 is protomer #1. An NNA Fortran program is run on the two protomers of the training BC, and then the resulting neighbor atoms are classified as H-bonds (Engh & Huber, 1991), VDW interactions (Bondi, 1964), or neither using yet another program. The four most dominant interface interactions involving four different atoms in each protomer are determined from the AAR. Both protomers are then transformed into double-centroid reduced representation (DCRR; Reyes, V.M. & Sheth, V.N., 2011) in preparation for the next step.

### 2.2.2 Step 2: Determination of the Interface Tetrahedral 3D SM Pair.

This step is the same as the determination of the 3D SM from the 3D binding site consensus motif in our previous work (Reyes, V.M., 2015a, c), except that instead of one 3D SM there are now two, one on the interface of each protomer. The general structure of the 3D SM pair at the interface of the two protomers in the binary complex is shown in Figure 3. The four centroids (backbone or side chain) corresponding to the four atoms engaged in interface interactions in each protomer are taken as forming the tetrahedral 3D SM in each protomer. The four centroids in each protomer are designated one root and three nodes in such a way that the root centroid in protomer #1 is associated with the root centroid in protomer #2, the node 1 centroid in protomer #1 is associated with the node 1 centroid in protomer #2, and so on, as depicted in Figure 3 (i.e., interactions R-R', $n_1$-$n_1$', $n_2$-$n_2$' and $n_3$-$n_3$').

The parameters of the pair of tetrahedral 3D SMs are then determined: the lengths of each side of the 2 tetrahedra (2x6 = 12 quantitative parameters), the distances between associated roots and nodes in the two tetrahedra (4 quantitative parameters), as well as the amino acid identities of the roots and nodes of the two tetrahedra (4x2 = 8 qualitative parameters), and their nature of interaction with the ligand, i.e., whether via backbone or side chain (4x2 = 8 qualitative parameters). The pair of tetrahedral search motifs from a binary complex therefore has 16 quantitative parameters and at least 16 qualitative parameters. Aside from the 3D SM parameters, the depth of burial of the 3D SM in the protomers was also determined using the CP and TS methods.

### 2.2.3 Step 3: Screening the Application Set for the Tetrahedral 3D SM Pair.

Each application structure is transformed into DCRR, and, using our screening algorithm, each is screened twice: once for the tetrahedral 3D SM from training structures for protomer #1, and a second time for the tetrahedral 3D SM from training structures for protomer #2. Application structures testing positive for the 3D SM of protomer #1 ("set 1"), and those testing positive for the 3D SM of protomer #2 ("set 2"), are the putative PPI partners for the particular BC in question. If set 1 has n members, and set 2 has m members, then there are nxm putative BCs between sets 1 and 2.

### 2.2.4 Step 4: Validation By Determining the 3D SM Burial Depth in Each Protomer.

The depth of burial of the 3D SM in each application structure testing positive above is then determined using CPM and TSM. Those whose CPi and/or TSi values are close enough to any of those in the training structures are selected and kept for the next step. The degree of similarity between the CPi and TSi values of the application and training structures are usually set arbitrarily, as every protein complex is different.

### 2.2.5 Step 5: Further Validation By Determining the Inter-Protomer Overlap in Complex.

IPO in the BCs may be performed in docked and undocked modes. The training structure complexes are naturally docked, so we determine the IPO in docked mode. The degree of overlap between the protomers when docked together can be determined by making use of the CP for either protomer (see Figure 5), as follows: Let plane P1: $A_1x + B_1y + C_1z = D_1$ be the equation of the CP for protomer #1; similarly, let plane P2: $A_2x + B_2y + C_2z = D_2$ be that for protomer #2. It is straightforward to determine the equation of the mid-plane Pm: $A_mx + B_my + C_mz = D_m$ that is equidistant from both CPs. Then the degree of IPO of protomer #1 into protomer #2 is the % of protomer #1 atoms on the side of plane Pm anterior to protomer #1, and conversely the degree of IPO of protomer #2 into protomer #1 is the % of protomer #2 atoms on the side of plane Pm anterior to protomer #2. Note that the CP for protomer #1 need not be parallel to that of protomer #2; even if they are not parallel, the mid-plane exists and be calculated from them

The application structures are undocked, thus the IPOs of the candidate PPI partners are determined in the undocked mode. If performed on undocked protomers, the equation of the cutting plane is first determined for each protomer, and the CPi of protomer #1 is taken as its overlap with its partner, protomer #2; similarly, the CPi of protomer #2 is taken as its overlap with its partner, protomer #1. Alternatively, the candidate PPI partners may first be docked together using steps outlined in the next section (2.2.6), and then the IPO determined in the docked mode as described in the previous paragraph. In either case, the putative complexes in the application set whose IPOs resemble those from their corresponding training structure are considered more likely candidates than those with IPOs that are quite different from the same.

### 2.2.6 Docking the Predicted Protein-Protein Interaction Partners Together.

PP docking is the juxtaposition of two protein molecules together in 3D space such that their orientations with respect to each other is specific, with specificity dictated by electrostatic and geometric complementarity at their interface, which in turn depends on the identities and rotameric conformations of the amino acids therein (Fernandez-Recio et al., 2002 & 2005; Smith & Sternberg, 2002). Docking candidate PPI partners together involves consideration of the 3D interface search motif tetrahedral pair (3D ISMTP; see Figure 3) constructed from the training structure for that particular BC as well as rotation and translation in 3D space so that two tetrahedra with four corresponding vertices are superimposed optimally, with optimality defined as the superimposed position of the two tetrahedra at which the root-mean-square deviation (RMSD) between the four corresponding vertices is minimum. Specifically, suppose protein A and protein B are predicted PPI partners, and the 3D SM for A are composed of root $R_a$, and nodes $n1_a$, $n2_a$ and $n3_a$; and correspondingly, those for B are $R_b$', $n1_b$', $n2_b$' and $n3_b$'. We must perform a rotation + translation of both proteins A and B such that (a.) $R_a$,, $n1_a$, $n2_a$ and $n3_a$ from protein A are superimposed optimally with R, n1, n2, and n3 in the 3D ISMTP, and (b.) $R_b$', $n1_b$', $n2_b$' and $n3_b$ from protein B are likewise superimposed optimally with R', n1', n2', and n3' in the 3D ISMTP. When these conditions are met, the resulting docked PPI partners is the structure of the putative BC that they are predicted to form.

## 3 Results.

The identities of the nine training structures corresponding to the nine binary complexes studied in this work are shown in Table 1. As for the identities of the application structures composed of 801 functionally unannotated but experimentally solved protein structures, please refer to Table 2 in Reyes, V.M. (2015c). Due to space limitations, we will describe the process in detail only for complex A, as the same process is applied to complexes B, C, D, H, I. P, Q and Z.

### 3.1 Determination of Protomer Interface Interactions from the Training Set.

The H-bond and VDW interactions between the two protomers in each BC training structure were determined by first performing between the two protomers, both in AAR at this stage, an NNA Fortran program we wrote. Atoms in one protomer close to atoms in the other protomer were then classified as HBs or VDW interactions based on the identities of the atoms involved, and the distances between them (Reyes, V.M. & Sheth, V.N., 2011; Reyes, V.M., 2015a,b). The HBs and VDW interactions at the interface in all nine training structures were determined using another Fortran program we wrote (data not shown). The four most dominant and/or recurrent in each case are selected by inspection, and allows us to come up with the 3D ISMTP in each case. The results are shown in Figure 7, parts 1 and 2, and, after transformation into DCRR, the distances between the corresponding interacting centroids are shown in Table 2, last four rows: RR', n1n1', n2n2' and n3n3' for all nine training structures,.

### 3.2 Determination of the Tetrahedral Interface Search Motif Pair.

It is first necessary to transform the training structures into DCRR in order to determine the 3D ISMTP between them. Once transformed into DCRR, the centroids corresponding to the four most dominant interactions determined above are taken as the root and nodes 1, 2 and 3 of the 3D ISMTP. For example, as in complex A, a Ser(172)-OG in protomer #1 is involved in H-bonding with the backbone NH of Arg(329) in protomer #2, then the side-chain centroid of Serine-172 is taken as a root or node in the 3D ISMTP for protomer #1, and the backbone centroid of Arginine-329 is taken as root or node in the tetrahedral 3D ISMTP of protomer #2.

The four intermolecular interactions that we determined to be most dominant between the two protomers in the nine training structures are shown schematically in Figure 6, parts 1 and 2. We use the notation R, n1, n2 and n3 for the root and three nodes of protomer #1, and R', n1', n2' and n3' for the root and three nodes of protomer #2. We note that their identities are shown in the first eight rows of Table 2. Meanwhile, the intermolecular interactions shown in Figure 3 comprise the last four rows of Table 2, i.e., RR’, n1n1’, n2n2’ and n3n3’. This now leaves us with the intramolecular distances between the root and three nodes of the tetrahedral 3D SM on each protomer: Rn1, Rn2, Rn3, n1n2, n1n3 and n2n3 for protomer #1, and R’n1’, R’n2’, R’n3’, n1’n2’, n1’n3’ and n2’n3’ for protomer #2. They are respectively shown on rows nine to 14 and rows 15 to 20 of Table 2, which are the six sides of the two interacting tetrahedra. Note that the illustration in Figure 3 shows the two interface 3D SMs as flat quadrilaterals instead of irregular 3D tetrahedra; this was done only for convenience; it must be emphasized that the 3D SMs must be thought of always as the latter.

**3.3 Validation: Positive and Negative Controls for the Interface 3D Search Motifs.**

The negative control structures (n=21) used for all protein families are shown in Table 3 of Reyes, V.M., 2015c. We used the same set of negative controls here as that for the six ligand binding sites studied in the aforementioned work. In all 18 cases (nine binary complexes, 18 protomers) the algorithm found no 3D SM in any of the negative control structures (data not shown). These results imply that the algorithm is specific for the respective BCs. As for positive control structures, we note that there are no other appropriate positive BC structures in the PDB as all of them have been used as training structures. We thus constructed artificial positive control structures from the negative control structures by replacing four appropriate amino acid residues in the latter to make a legitimate 3D SM for the particular BC protomers. These artificially mutated structures are then screened for the appropriate 3D SM using our screening algorithm. In all cases, the algorithm detected the artificially embedded 3D SM for the particular BC (data not shown). These results imply that the screening algorithm is indeed sensitive for the terahedral 3D SM corresponding to the particular monomer in the training BC.

**3.4 Screening the Application Set for the Tetrahedral Interface Search Motif Pair.**

The 801 FUA protein application structures were screened for each tetrahedral 3D SM using the screening algorithm we developed earlier for PLIs (Reyes, V.M. & Sheth, V.N., 2011; Reyes, V.M., 2015a). The percentage of structures in the application set that tested positive for the 3D SM of either protomer in the nine training BCs are shown in Table 3. Note that in some of these structures, more than one BS have been detected, and that the interacting residues in each of the above protomer pairs are known specifically.

Due to space limitations, we cannot show the identities of all these positive application structures, so we shall show the details of the results for each step in the screening process only for complex A. For the other complexes, we shall only show the final results. For all complexes, the details of the results can be made available upon request.

**3.5 Using the CPM and TSM to Determine Ligand Burial Depth in Each Protomer.**

After detecting the appropriate protomer 3D SM in some of the application structures, they were tested for the depth of burial of these 3D SMs using the CPM and TSM so that the values may be compared to those of the training structures. First, those with CPi or TSi similar to those of the training structure were sequestered, then from these, those with both CPi and TSi similar to those of the training structure were selected. UNIX scripts were used to perform these selections from the output of the Fortran programs used in the 3D SM detection in the earlier stage of the screening process.

Table 4, Panels A and B (for complex A), and Table 5 (for complexes B-Z) show the structures that tested positive after the third step of our elimination process (refer to Figure 4). The identities of the amino acids, including residue number, comprising the root and three nodes in each positive application structure for all nine BCs are shown. Note that for complex A, there are 11 candidate protomer #1 and 3 candidate protomer #2 proteins. Thus 11 x 3 = 33 putative complexes A are possible from the 801 application structures. For complex B, 11 candidate protomer #1 and 27 candidate protomer #2 proteins result in 11 x 27 = 297 possible complexes B. For complex C, 2 candidate protomer #1 and 31 candidate protomer #2 proteins result in 2 x 31 = 62 possible complexes C. For complex D, 7 candidate protomer #1 and 9 candidate protomer #2 proteins result in 7 x 9 = 63 possible complexes D. For complex H, 8 candidate protomer #1 and 15 candidate protomer #2 proteins result in 8 x 15 = 120 possible complexes H. As for complex I, no protomer #1 was detected, so even though 37 protomer #2 were detected, no possible complex I are possible. For complex P, 9 candidate protomer #1 and 12 candidate protomer #2 proteins result in 9 x 12 = 108 possible complexes P. For complex Q, 2 candidate protomer #1 and 8 candidate protomer #2 proteins result in 2 x 8 = 16 possible complexes Q. And finally for complex Z, 9 candidate protomer #1 and 56 candidate protomer #2 proteins result in 9 x 56 = 504 possible complexes Z.

### 3.6 Using the "Cutting Plane" Method to Determine Protomer Overlap.

The results we obtained for complex A is shown in Table 4, Panel C. Here, 12 structures testing positive for protomer #1 and 3 structures testing positive for protomer #2 resulted in a total of 36 possible putative binary complexes corresponding to complex A. For comparison, the IPOs between the protomers of the training set structure were also determined using the same method. The values we obtained from the training structure were 1.54% and 9.27%, respectively. Since the training structure is a docked complex (i.e., it is an experimentally solved complex structure), we must give some marginal leeway when comparing the IPO results from the application structures, which were determined using undocked structures.

The results shown in Table 4, Panel C, show that (1.) 1nkv:C with root at D-94 and 1o69:A with root at T-328, with overlaps of 5.88% and 9.64%, respectively; (2.) 1zsw:A with root at D-187 and 1o69:A with root at T-328, with overlaps of 2.45% and 9.64%, respectively; (3.) 1nkv:C with root at D-94 and 2a2o:F with root at T-13, with overlaps of 5.88% and 10.24%, respectively; and (4.) 1zsw:A with root at D-187 and 2a2o:F with root at T-13, with overlaps of 2.45% and 10.24%, respectively, resemble the training structure much better than the rest of the putative complexes do, and as such they are deemed to be more probably true positives then the rest. The overall final results for all eight complexes are shown in Table 6; note that no final positive structure was found for monomer 1 of complex I (although some final positive structures were found for monomer 2), hence there are no candidates for complex I.

## 4 Discussion.

We have developed a procedure for predicting PPI partners based on an algorithm that searches for tetrahedral configurations of amino acid backbone or side chain centroids in protein 3D structures. We call these configurations 3D SMs when they are alone, and 3D ISMTP when docked together in the BC. These are determined from a training structure for the BC under study. A set of application structures, which are proteins whose 3D structures have been solved but whose function(s) are unknown, are then searched for the existence of these configurations using a multi-step screening procedure we developed previously. Structures testing positive for either tetrahedron are candidate partners, which are then validated by determining the degree of 3D SM burial in each protomer, and then the physical overlap between the two protomers docked together as the putative BC.

### 4.1 Notable Features of the Method.

The search algorithm on which our method is based was first used to search for LBSs in protein structures (Reyes, V.M. & Sheth, V.N., 2011; Reyes, V.M., 2015a, c). Here we extend the procedure to the prediction of PPI partners. The screening algorithm, however, is the same: it searches for tetrahedral configurations of points corresponding to the amino acid residue backbone or side chain centroids located on the complexation interface of a protein protomer and interacting with a corresponding tetrahedron on the other protomer of the complex. This method is also a semi-docking method, since residues in protomer #1 interacting with residues in protomer #2 are all specifically known, thus the structures of the putative protomer partners can be translated and rotated to superimpose on the 3D ISMTP using RMSD minimization.

The search algorithm on which our method is based is analytical, and this quality distinguishes it from most currently available methods, such as SVM and neural networks, which are stochastic in nature. Its analytical nature gives it more objectivity than other methods. Advantageous features of the previous work on LBS screening (Reyes, V.M., 2015a, c) should also apply here. Other notable features of this method are: the due consideration given to VDW interactions besides HB interactions; the due consideration given to backbone interactions besides side-chain interactions; its extensibility to other BCs as well as to higher-order complexes; its amenability to high-throughput implementation; its high specificity to the BC being screened for; its use of fuzzy factor and reduced protein representation (i.e., DCRR) to counteract the effects of protein flexibility and dynamics; and the use of a novel interfacial tetrahedron pair data structure. Finally, this method allows function prediction, as it is specific for the PPI partners under consideration. This main advantage of the multi-step elimination procedure is it can be automated and ran in batch (high-throughput) mode, without the requirement for human intervention, a feature desired of analytical tools for large datasets.

For all nine BCs treated in this study, there is only one training structure. We have shown earlier that a single training structure works for our screening algorithm (V. M. Reyes, 2015c). In such instances, the screening results may not be as sensitive as when there are multiple training strictures, but the specificity is usually enhanced. This is to be expected since a singleton training structure may considered a "pure" set.

The values we obtained for degrees of protomer overlap (i.e., IPO) in the candidate complexes are worth noting. Those with high degrees of overlap suggest that the protomers might be required to undergo large conformational changes upon complex formation with its partner. Conversely, those with smaller degrees of overlap are more likely to bind their partners without the need for drastic conformational changes. This is akin to the case of a deeply buried LBS versus a more exposed one. In the former, the receptor protein most likely undergoes a drastic conformational change upon ligand binding as compared to the latter. Therefore the depth of burial of the 3D SM in a protomer as well as the degree of overlap (as given by the IPO) gives us a clue about the flexibility and dynamics of the proteins involved.

Finally, our method is novel and analytical (deterministic, exact), specific for the PPI partners being predicted, and fully automatable and thus amenable to high-throughput implementation.

### 4.2 Limitations of the Method.

Our procedure requires that there be at least one solved structure of the BC being studied for use as a training structure. This is certainly a limitation of the method since experimental determination of protein complex structures are currently feasible only by x-ray crystallography (e.g., it is still problematic via protein NMR). In addition, solving protein complex structures by x-ray crystallography is technically more challenging (especially of higher-order complexes than binary ones) than solving that of single-chain proteins. As it currently stands, our procedure has been worked out only in the case of BCs. In the near future, we shall extend our procedure to the prediction of higher-order complex partners. It will involve similar procedures, but computational demands will be exponentially higher since the protomer proteins in the n-ary complex will have to be dealt with two at a time, requiring an additional factor of C(n,2) to the computational demands.

If the protomers in the binary complex undergo large conformational changes upon binding each other, then our tetrahedral 3D SM method is unlikely to work. This is because the relative geometries of the interacting protein residues in the two undocked protomers (which are the ones used in screening) are likely to differ drastically from their relative geometries in their corresponding docked configurations (which is the case in the application structures). The incorporation a fuzzy factor and the use of reduced representation of the proteins involved (i.e., DCRR) do counteract this effect to some extent, but if the conformational changes are too drastic, these measures will not suffice. Unfortunately, there is not much that can be done regarding drastic conformational changes, as there is currently no effective way to predict protein dynamics except to employ complex molecular dynamics simulations. The protein dynamics problem remains largely unsolved. Perhaps a form of solution will come from protein NMR spectroscopy.

## 5 Summary and Conclusion.

We have developed a novel analytical method for the prediction of specific PPI partners. The method is currently designed for BCs, but will be extended to higher-order complexes in our future work. The method has been applied to a set of some 801 experimentally solved protein structures in the PDB which currently do not have functional annotations. The method is a semi-docking procedure, allowing the candidate PPI partners to be docked together in a complex. The method is an extension of the tetrahedral 3D SM-based methodology previously developed by us for the prediction of LBSs in proteins. Since the method is highly specific with respect to the PPI partners predicted, it is a highly viable tool for protein function prediction.

**Acknowledgement.** The author thanks the San Diego Supercomputer Center, the UCSD Academic Computing Services, and the UCSD Biomedical Library for the use of their various computing and other resources. He is also grateful to Professors Laurence Brunton, Dept. of Pharmacology, UCSD, and Robert Pozos, Dept. of Biology, SDSU, for their support. This work was supported by an Institutional Research and Academic Career Development Award to the author, NIGMS, NIH grant GM 68524.

**References.**

**Figure Legends.**

**Figure 1. Interactions at the Interface of the Two Protomer Proteins in a Binary Complex.** Two proteins, protomer #1 (left; lavender surface) and protomer #2 (right; brown surface), are shown interacting at their mutual interface. The four small red balls represent the side chain or backbone centroids of four amino acid residues on the protomer #1 interface that interact with four of the same on the protomer #2 interface, represented by the four small blue spheres. The interactions between the small red and blue spheres are either HBs or VDW attractive forces, and are responsible for the formation and stabilization of the BC between the two protomers. The four small red spheres form a tetrahedron, so do the four small blue ones. These two tetrahedra each form a 3D SM, similar to the one we presented previously in the case of PLIs (Reyes, V.M. & Sheth, V.N., 2011; Reyes, V.M., 2015a, b). Each can then be searched for in protein structures using the same search algorithm we reported previously (ibid.).

**Figure 2. Our Overall Procedure for Detecting Protein-Protein Interaction Partners.** A schematic diagram of the procedure for the prediction of PPI partners is shown. An experimental 3D structure of the BC being studied is required, as the interface 3D SM pair is derived from it. Using our screening algorithm, the application structures are screened twice: once for the protomer #1 interface motif, and a second time for the protomer #2 interface motif. Structures testing positive for the former, and those testing positive for the latter, are candidate (putative) PPI partners.

**Figure 3. Structure of the Interface 3D Search Motif Tetrahedra Pair.** The 3D ISMTP is shown detached from its source BC (please refer to Figure 1). It is actually a pair of interacting tetrahedra, since each individual 3D SM is a tetrahedron. In the diagram each is shown as a flat rectangle for simplicity. Interactions between the two protomers are represented by RR', n1n1', n2n2' and n3n3' (green broken lines), where primed and unprimed letters refer to corresponding tetrahedral vertices.

**Figure 4. The Multi-Step Elimination Process.** The 801 FUA application structures were first screened for the 3D SM determined from the protomers in the nine BC training structures. Those testing positive were then subjected to the CPM and TSM to determine the depth of burial of the 3D SM(s) found on them. Those with 3D SMs with the same depth of burial as those in the training structures were then selected. These smaller group was then analyzed as to the degree of overlap between the protomer partners when docked as the BC, with those having IPO values similar to the corresponding training structure selected over the others.

**Figure 5. Extension of the Cutting Plane and Tangent Sphere Methods to Protein-Protein Complexes.** We originally devised the CPM and TSM for PLIs, but it can be easily extended to PPIs, as illustrated here. Consider a BC consisting of protomers #1 (solid red line) and #2 (solid green line). At their interface is the pair of interacting 3D SMs as described in the text and shown here as a red quadrilateral (on protomer #1) and a green one (on protomer #2). The degree of overlap between the protomers when docked together can be approximated by determining the mid-plane between the CPs of protomers #1 and #2, as described in the text.

**Figure 6. The Four Most Dominant Intermolecular HB and VDW Interactions at the Interfaces of the Nine Training BCs.** The nine panels of this figure show the four most dominant interface interactions between the two protomers in the nine BCs that we used in building the 3D ISMTP. The tetrahedral 3D SM in each protomer is depicted by the tetragon with four tan broken-line sides near the interface of the BC. The intermolecular interactions between protomers are depicted by solid red and blue lines. The identities of the amino acids comprising the 3D ISMTP are identified, as well as whether each refers to a sidechain centroid, "(s)", or a backbone centroid, "(b)".

**Table Legends.**

**Table 1. The Training Structures.** The first column shows the PDB ID of the binary complex under study, as well as the chain IDs of the component protomers (the first chain ID shown is protomer #1; the second, protomer #2). The second column shows the one-letter abbreviations we used in the text to identify the complexes. The last column identifies these binary complexes.

**Table 2. Interface 3D SM Pair Parameters.** The 32 parameters of each 3D ISMTP for each training BC are shown. Each component tetrahedron has 14 parameters (8 qualitative and 6 quantitative), and in addition, there are four intervertex sides (AA', BB', CC' and DD') representing interactions between the two component protomers. That makes a total of 32 parameters – 16 of which are quantitative (lengths of sides) and 16 are qualitative (identities of the amino acids and nature of interaction: backbone or side chain).

**Table 3. 3D SM Screening Results.** The first column shows the two-letter abbreviations we used in the text to identify the protomer components of the BCs; the letter part is identical to the ones in Figure 1 (second column); the number part refers to the protomer component. Since the same application set of 801 FUA proteins structures (see text) was used in screening for the 9 complex search motifs, the percentages given in the last column are all based on this number.

**Table 4. Complete Screening Data for Complex A.** Panel A shows depth of burial from CPM and TSM of the 3D SM in the putative protomers #1 and #2, as well as those of the lone training structure. Panel B shows the identities of the best candidate structures for protomers #1 and #2, along with their four nodes and chain IDs. Panel C shows the determination of overlap between protomers #1 and #2 for all possible binary complexes from the candidate protomers, and how the best candidate BC is selected based on these information.

**Table 5. Panels A-H. Detected Interface 3D SMs and the CPM and TSM Indices.** In this set of tables, the first column of each component table shows the PDB ID of the application structure testing positive, while the second, third, fourth and fifth columns show the "root", "node1", "node2" and "node3" of the detected 3D SM; finally, the sixth column shows the chain ID of the structure where the 3D SM was detected. Incorporation of the CPi and TSi information allows one to further narrow down the putative positive structures (see Figure 4).

**Table 6. Application Structures that Tested Positive.** The structures from the set of 801 FUA proteins (Table 2) in the PDB that tested positive of the 3D SMs of protomers #1 and #2 in the nine BCs (see Table 1) are summarized in this table. Note that most of the structures are still functionally unannotated at the time of this writing, as shown by the scarcity of entries in the last column, which is the published reference papers for the particular structure.

# FIGURES:

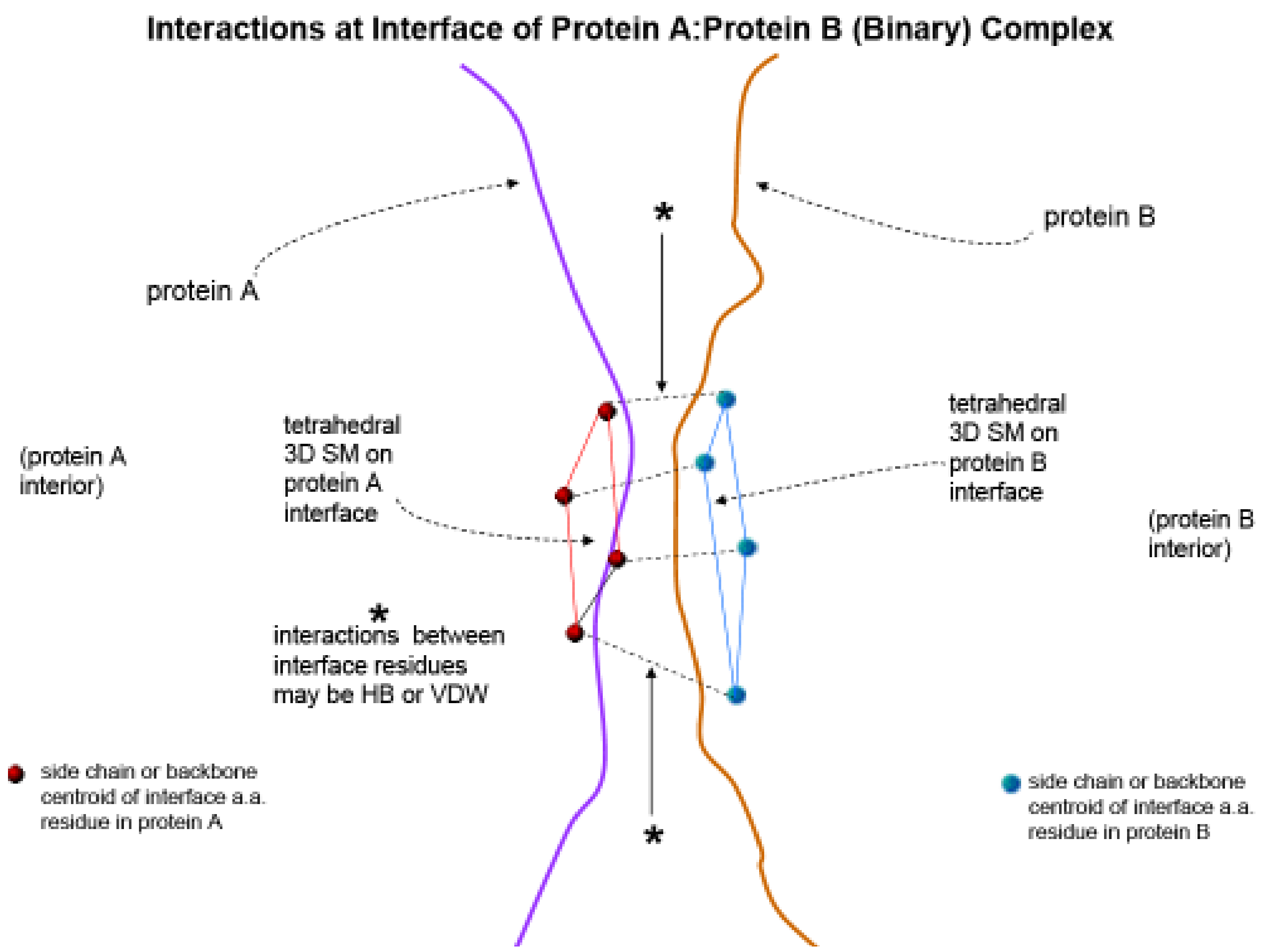

**Figure 1.**

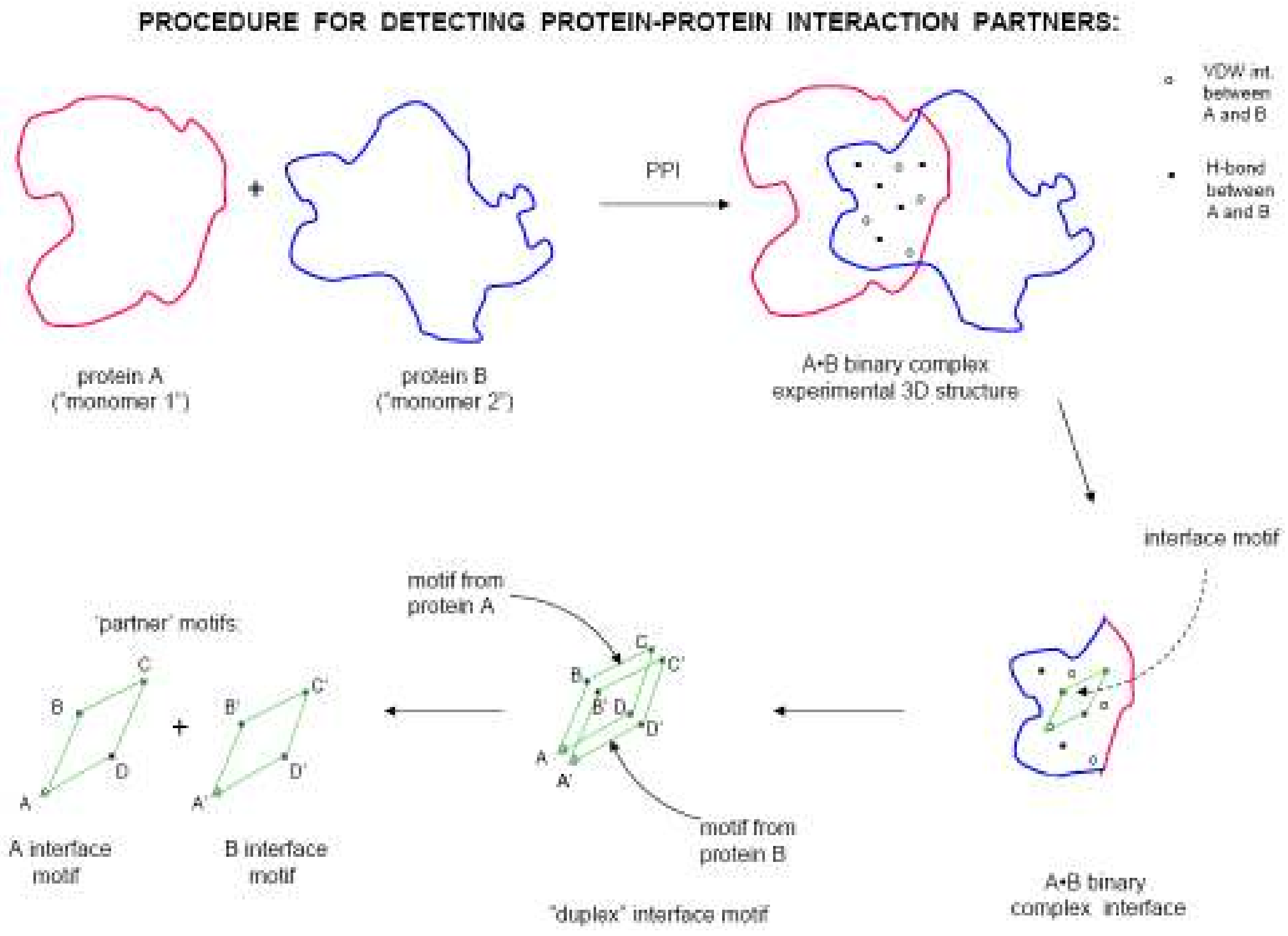
PROCEDURE FOR DETECTING PROTEIN-PROTEIN INTERACTION PARTNERS:
VDW int. between A and B
H-bond between A and B
PPI
protein A ("monomer 1")
protein B ("monomer 2")
A•B binary complex experimental 3D structure
interface motif
A•B binary complex interface
motif from protein A
motif from protein B
"duplex" interface motif
'partner' motifs:
A interface motif
B interface motif
A B C D
A' B' C' D'

**Figure 2.**

INTERFACE TETRAHEDRAL SEARCH MOTIF PAIR:

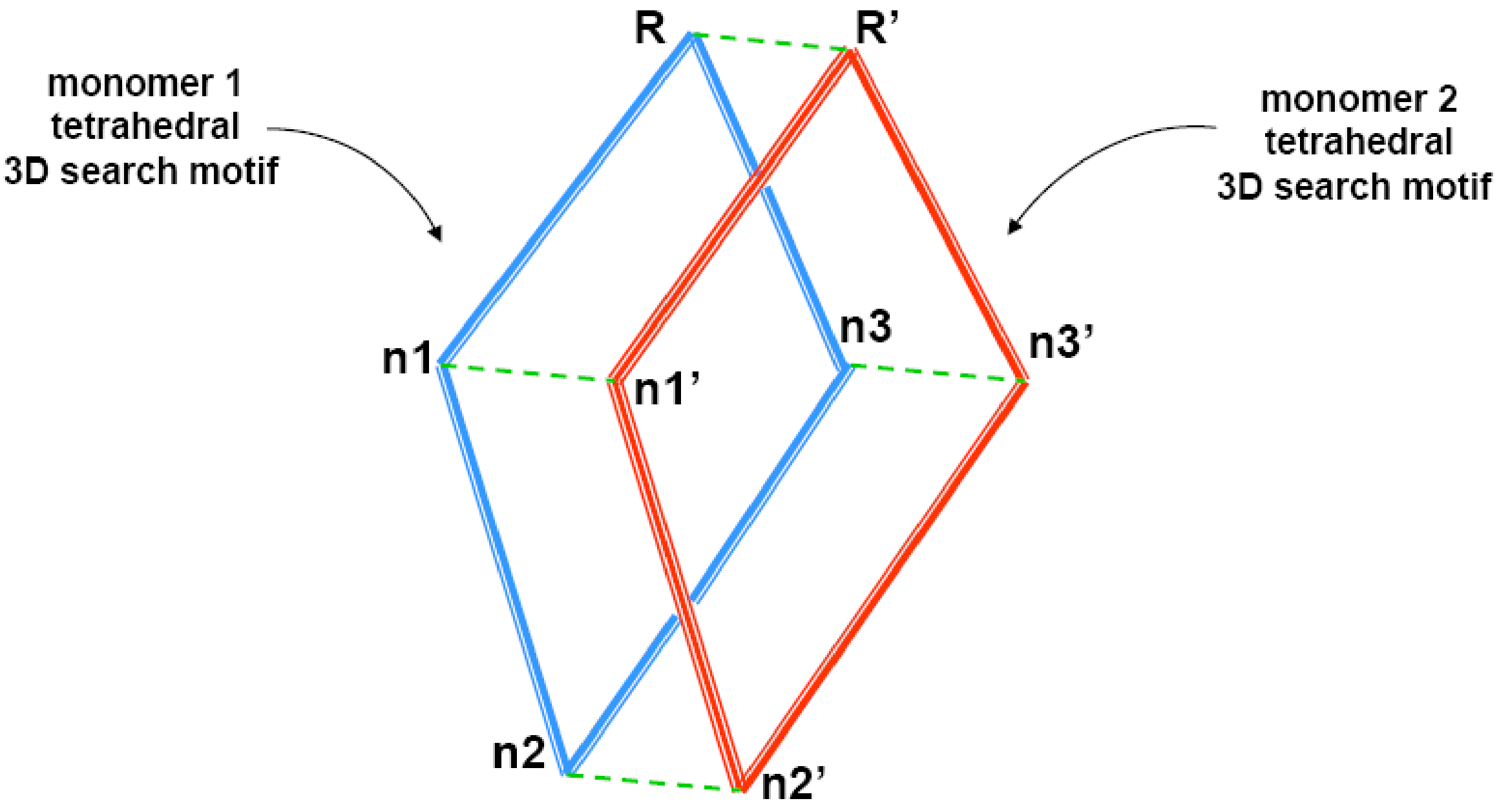

**Figure 3.**

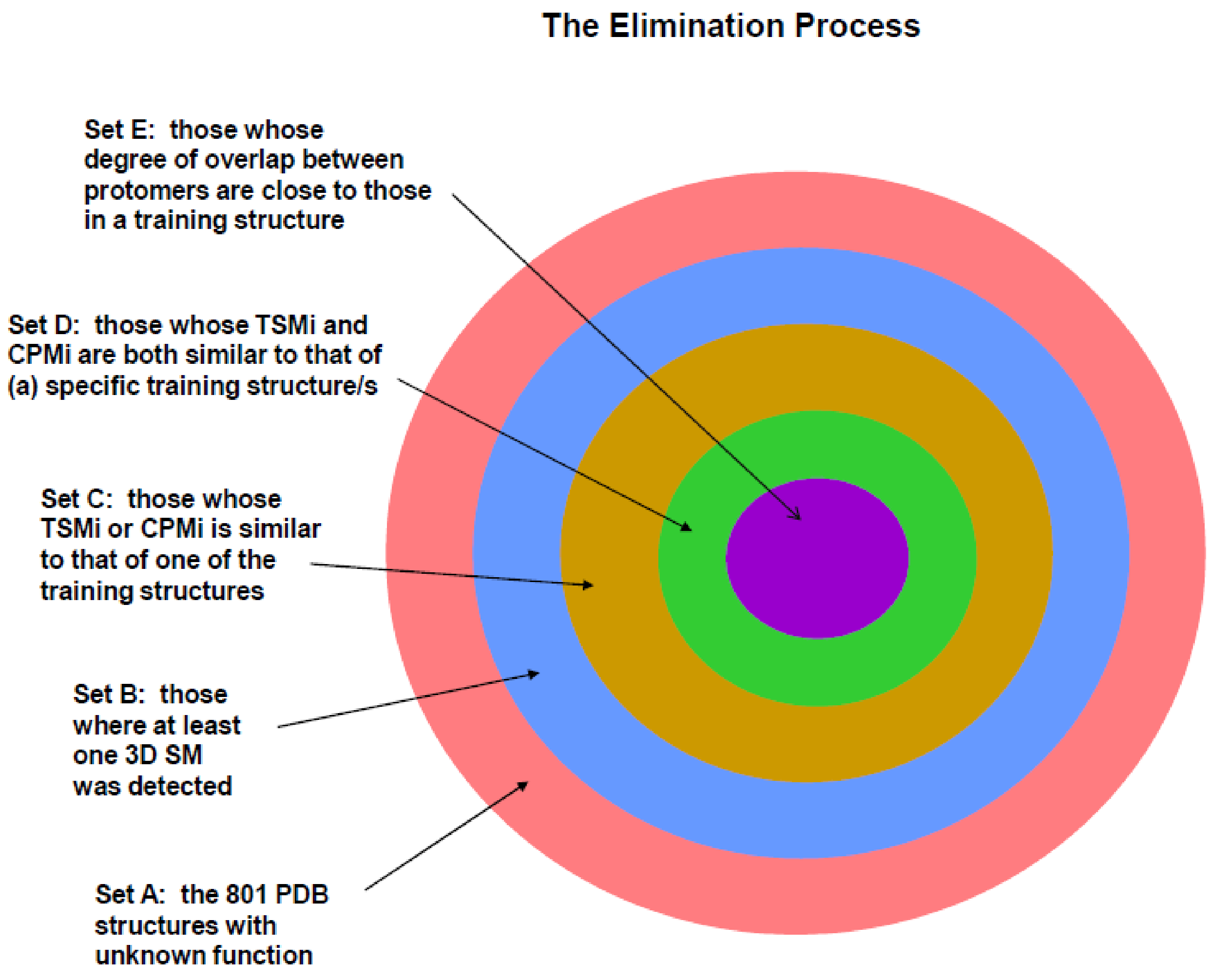
The Elimination Process
Set E: those whose degree of overlap between protomers are close to those in a training structure
Set D: those whose TSMi and CPMi are both similar to that of (a) specific training structure/s
Set C: those whose TSMi or CPMi is similar to that of one of the training structures
Set B: those where at least one 3D SM was detected
Set A: the 801 PDB structures with unknown function

**Figure 4:**

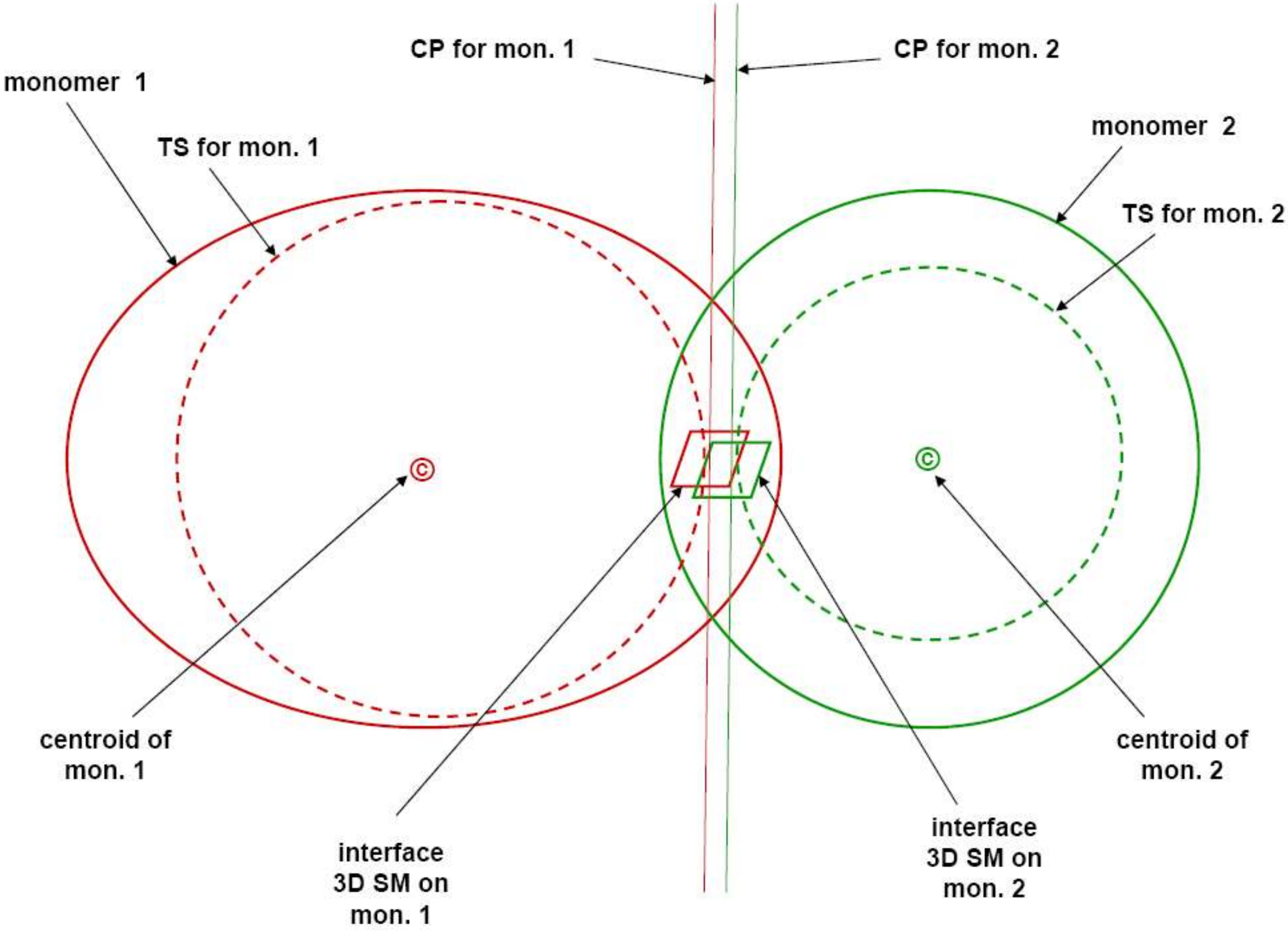
CP for mon. 1
CP for mon. 2
monomer 1
TS for mon. 1
monomer 2
TS for mon. 2
centroid of
mon. 1
centroid of
mon. 2
interface
3D SM on
mon. 1
interface
3D SM on
mon. 2

**Figure 5.**

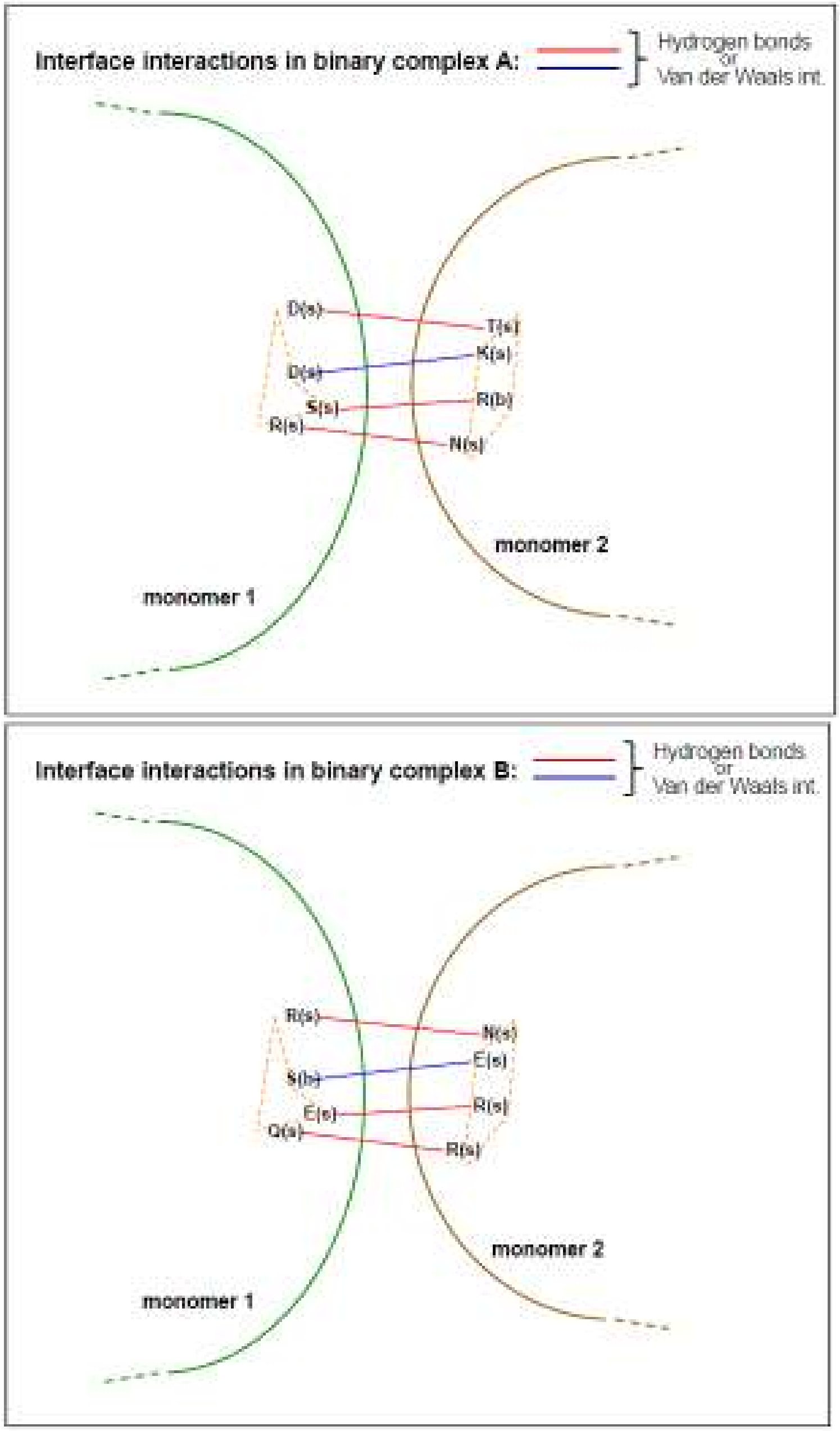

**FIGURE 6 (part 1 of 5)**

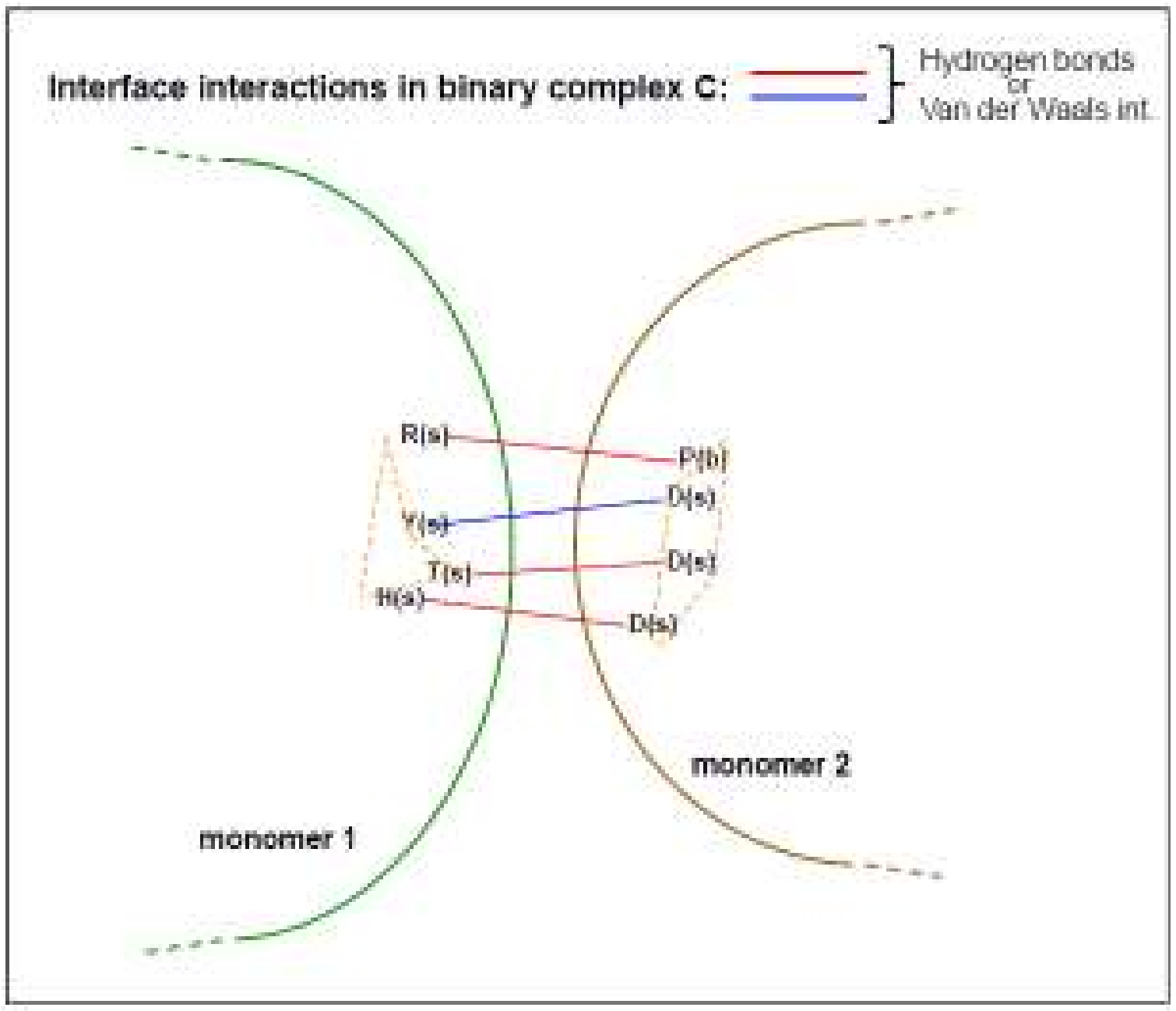

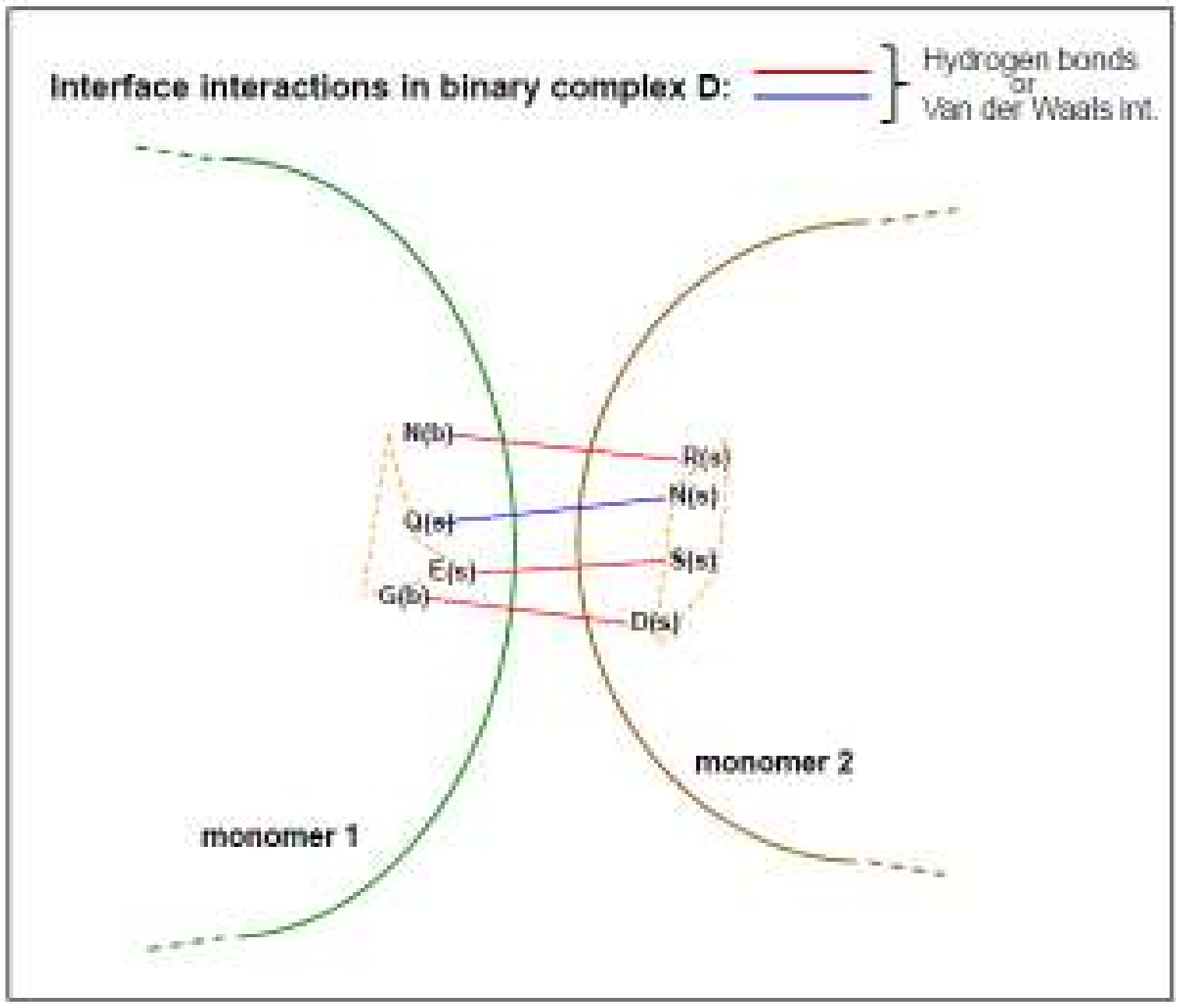

**FIGURE 6 (part 2 of 5)**

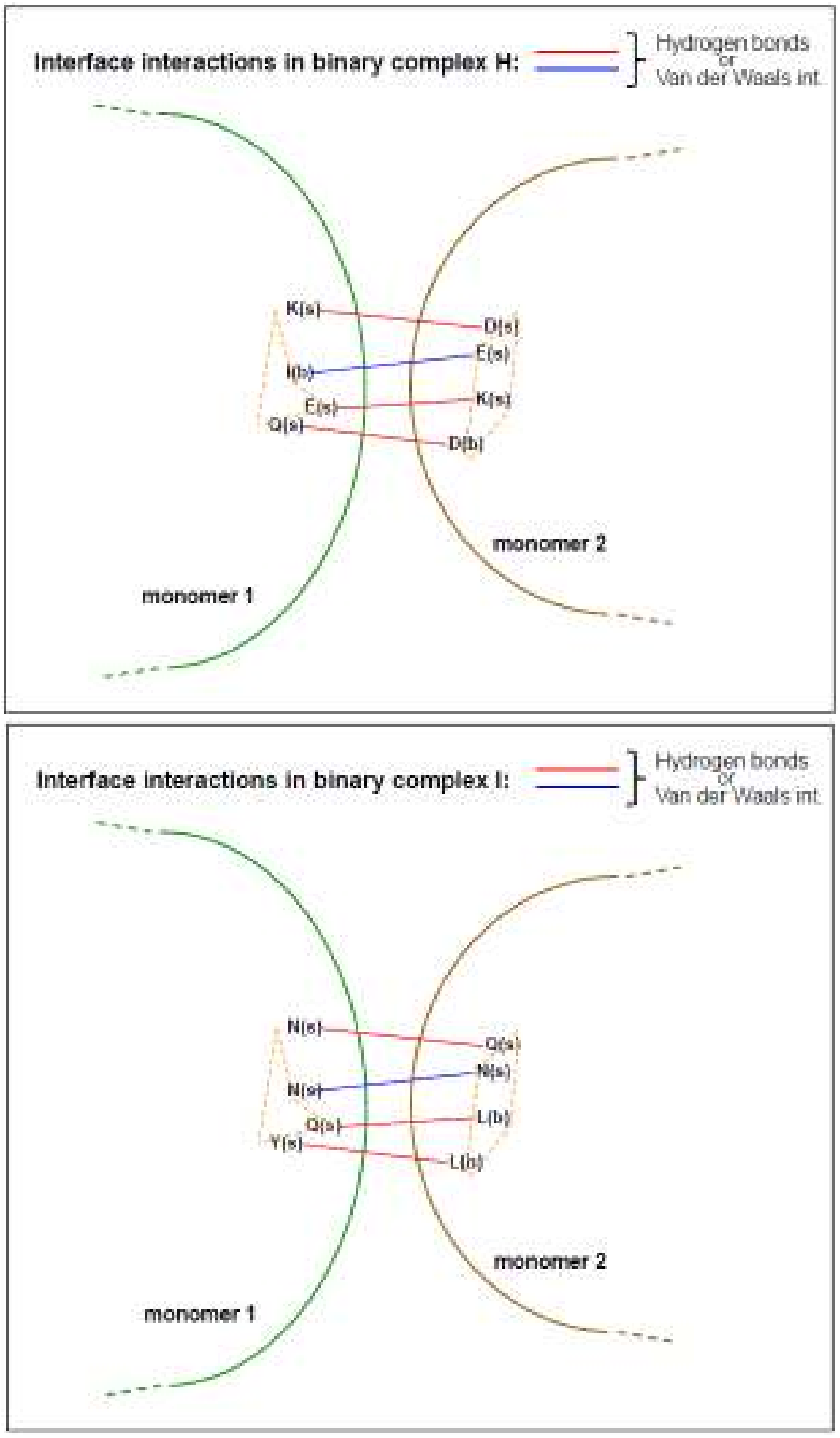

**FIGURE 6 (part 3 of 5)**

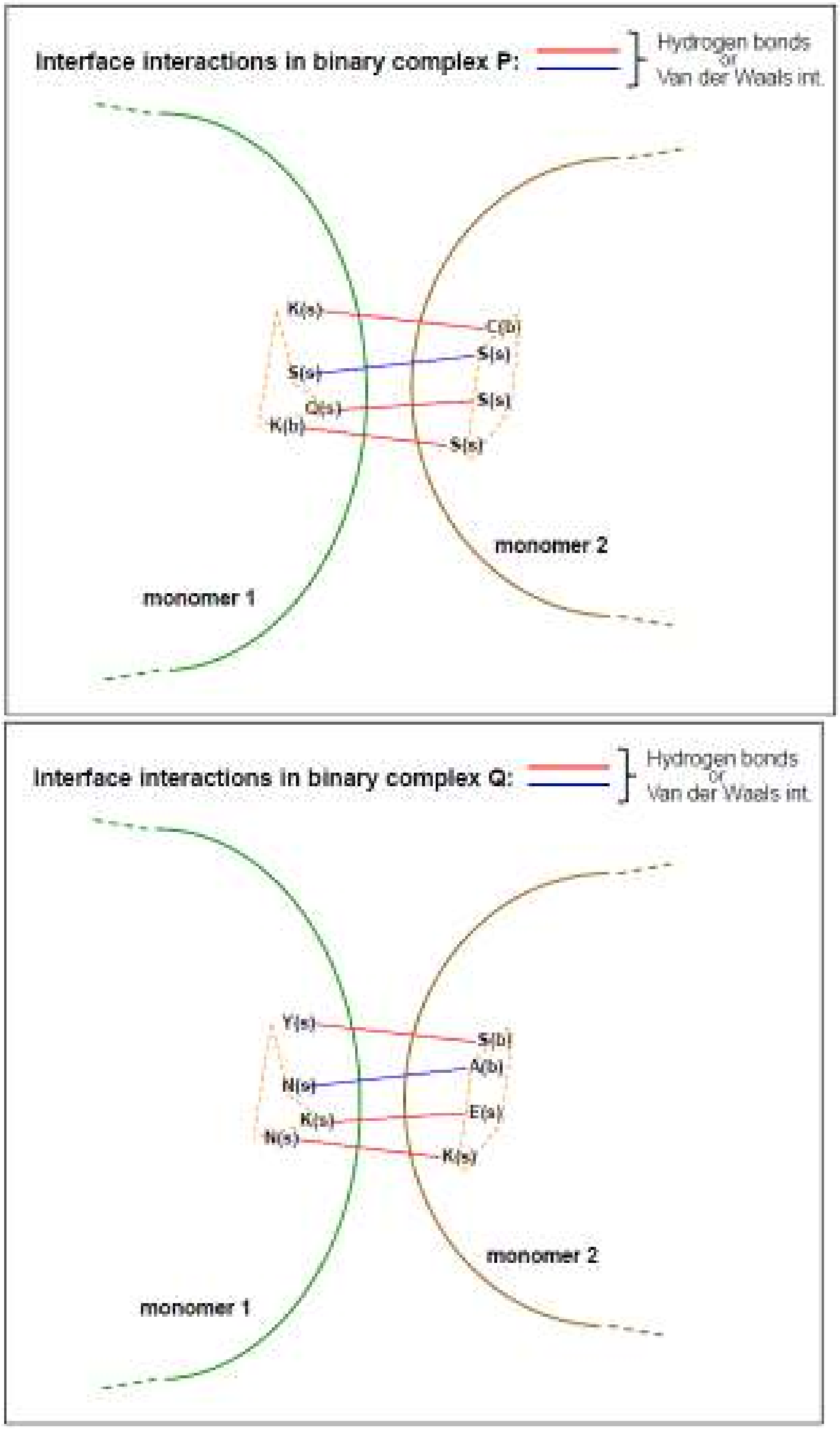

**FIGURE 6 (part 4 of 5)**

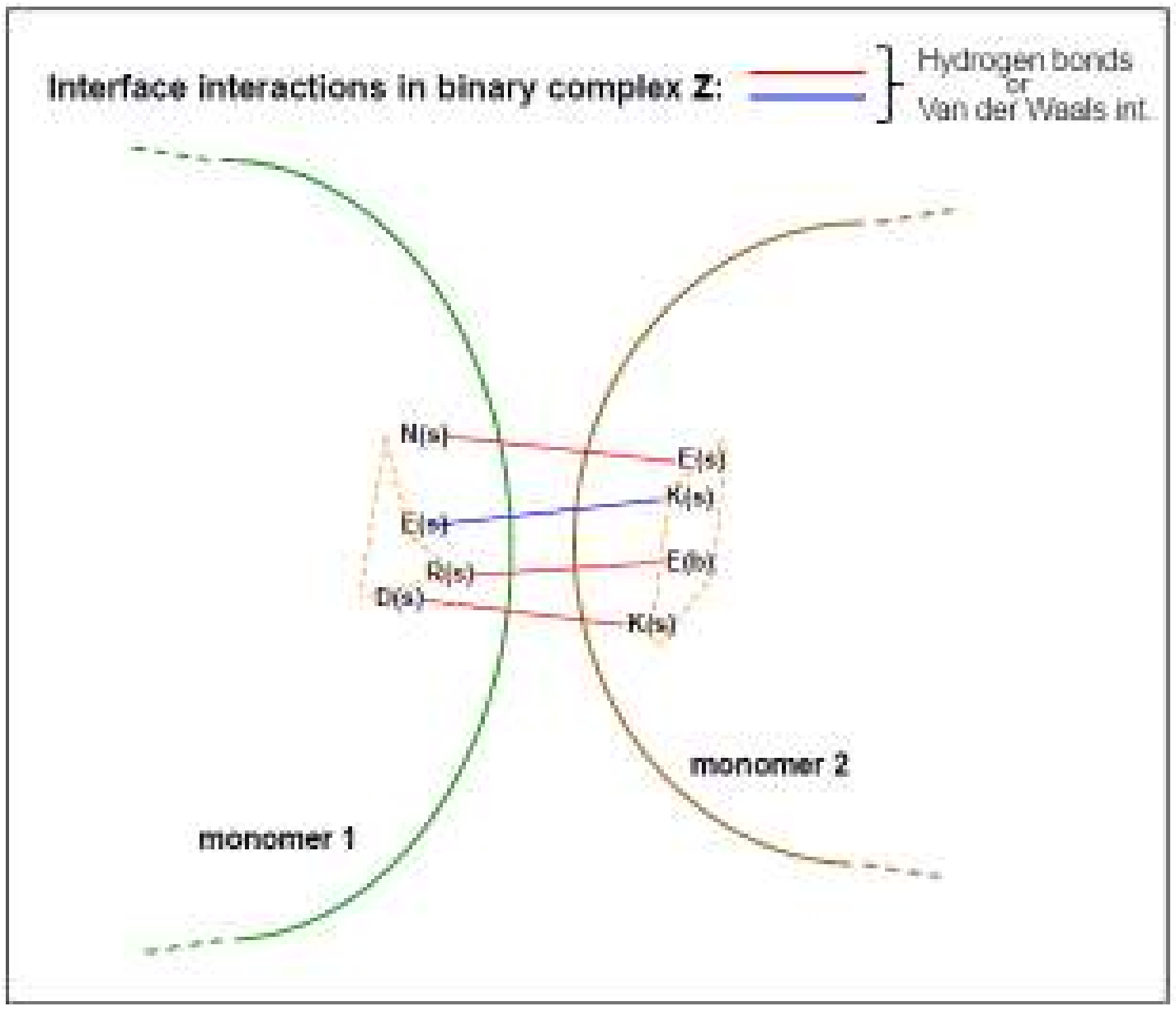

**FIGURE 6 (part 5 of 5)**

# TABLES:

## The Training Structures

| | | |
|---|---|---|
| **Complex A:** | **1C1Y:A,B** | **RAP•Gmppnp complexed w/ c-RAF1 Ras-binding domain (RAFRBD)** |
| molecule 1 | *Homo sapiens* | Ras-related protein RAP-1A (fragment, residues 1-167) with bound Gmppnp |
| molecule 2 | *Homo sapiens* | Proto-oncogene ser/thr protein kinase RAF-1 (fragment RAFRBD, residues 51-131) |
| **Complex B:** | **1CXZ:A,B** | **RHOA complexed w/ protein kinase PKN/PRK1 effector domain** |
| molecule 1 | *Homo sapiens* | His-tagged transforming protein RHOA (resdiues 1-181) |
| molecule 2 | *Homo sapiens* | effector domain of PKN (fragment, residues 13-98) |
| **Complex C:** | **1DS6:A,B** | **RAC complexed w/ RHOGD1** |
| molecule 1 | *Homo sapiens* | Ras-related C3 botulinum toxin substrate 2 (syn.: P21-RAC2) |
| molecule 2 | *Homo sapiens* | Rho GDP-dissociation inhibitor 2 (syn.: rho GDI 2, rho-GDI beta) |
| **Complex D:** | **1E96:A,B** | **RAC complexed w/ P67PHOX** |
| molecule 1 | *Homo sapiens* | Ras-related C3 botulinum toxin substrate 1 (syn.: RAC1) |
| molecule 2 | *Homo sapiens* | neutrophil cytosol factor 2 (NCF-2) TPR domain, residues 1-203 (syn.: P67PHOX) |
| **Complex H:** | **1FQ1:A,B** | **kinase-associated phosphatase (KAP) complexed w/ phospho-CDK2** |
| molecule 1 | *Homo sapiens* | cyclin-dependent kinase inhibitor 3 (E.C.3.1.3.48) |
| molecule 2 | *Homo sapiens* | cell division protein kinase 2 (E.C.2.7.1.-) |
| **Complex I:** | **1FC2:C,D** | **immunoglobulin Fc complexed w/ fragment B of protein A** |
| molecule 1 | *Homo sapiens* | immunoglobulin Fc fragment |
| molecule 2 | *Staphylococcus aureus* | fragment B of protein A |
| **Complex P:** | **1MCO:L,H** | **Ig G1 w/ a hinge deletion, light chain•heavy chain complex** |
| molecule 1 | *Homo sapiens* | immunoglobulin G1 (IgG1) with a hinge deletion, light chain |
| molecule 2 | *Homo sapiens* | immunoglobulin G1 (IgG1) with a hinge deletion, heavy chain |
| **Complex Q:** | **1JDH:A,B** | **beta-catenin complexed w/ HTCF-4** |
| molecule 1 | *Homo sapiens* | beta-catenin (residues 135-663) |
| molecule 2 | *Homo sapiens* | transcription factor TCF-4 (residues 12-49) |
| **Complex Z:** | **3INK:A,B** | **interleukin-2 homodimer** |
| molecule 1 | *Homo sapiens* | Interleukin-2 Cys125Ala mutant |
| molecule 2 | *Homo sapiens* | Interleukin-2 Cys125Ala mutant |

**Table 1.**

## Interface 3D Search Motif Pair Parameters

| | Binary Complex (one-letter abbrev.) | | | | | | | | |
|---|---|---|---|---|---|---|---|---|---|
| | A | B | C | D | H | I | P | Q | Z |
| R | D(s) | R(s) | R(s) | N(b) | K(s) | N(s) | K(s) | Y(s) | N(S) |
| n1 | D(s) | S(b) | Y(s) | Q(s) | I(b) | N(s) | S(s) | N(s) | E(S) |
| n2 | S(s) | E(s) | T(s) | E(s) | E(s) | Q(s) | Q(s) | K(s) | R(S) |
| n3 | R(s) | Q(s) | H(s) | G(b) | Q(s) | Y(s) | K(b) | N(s) | D(S) |
| R' | T(s) | N(s) | P(b) | R(s) | D(s) | Q(s) | C(b) | S(b) | E(S) |
| n1' | K(s) | E(s) | D(s) | N(s) | E(s) | N(s) | S(s) | A(b) | K(S) |
| n2' | R(b) | R(s) | D(s) | S(s) | K(s) | L(b) | S(s) | E(s) | E(b) |
| n3' | N(s) | R(s) | D(s) | D(s) | D(b) | L(b) | S(s) | K(s) | K(S) |
| Rn1 | 9.038 | 23.715 | 11.679 | 5.464 | 5.602 | 18.664 | 14.460 | 39.529 | 13.163 |
| Rn2 | 6.898 | 5.889 | 19.901 | 15.631 | 10.855 | 16.387 | 7.666 | 8.877 | 10.009 |
| Rn3 | 13.945 | 15.515 | 11.298 | 12.442 | 11.814 | 15.900 | 14.378 | 19.548 | 5.853 |
| n1n2 | 15.928 | 28.329 | 8.596 | 18.923 | 10.129 | 8.113 | 7.380 | 36.818 | 13.424 |
| n1n3 | 22.579 | 8.784 | 18.771 | 14.339 | 8.854 | 7.869 | 26.682 | 56.363 | 12.250 |
| n2n3 | 7.923 | 19.867 | 24.554 | 6.038 | 18.362 | 8.113 | 19.487 | 23.612 | 4.181 |
| R'n1' | 9.619 | 26.132 | 17.876 | 6.453 | 8.306 | 15.809 | 16.887 | 38.792 | 12.738 |
| R'n2' | 4.586 | 6.767 | 21.902 | 15.374 | 7.371 | 10.667 | 7.577 | 8.296 | 11.537 |
| R'n3' | 13.687 | 16.292 | 7.636 | 11.068 | 16.069 | 16.959 | 15.196 | 23.072 | 5.180 |
| n1'n2' | 13.741 | 30.830 | 5.473 | 19.532 | 13.395 | 10.411 | 10.646 | 34.595 | 15.079 |
| n1'n3' | 19.938 | 10.534 | 23.014 | 15.996 | 8.303 | 7.152 | 27.306 | 53.150 | 13.930 |
| n2'n3' | 10.401 | 21.092 | 26.023 | 4.720 | 19.143 | 10.062 | 17.289 | 22.746 | 7.420 |
| RR' | 4.516 | 5.944 | 4.932 | 5.861 | 3.761 | 4.498 | 4.317 | 6.293 | 4.313 |
| n1n1' | 5.061 | 4.571 | 5.867 | 5.060 | 5.218 | 4.777 | 3.469 | 4.580 | 5.523 |
| n2n2' | 3.956 | 6.133 | 4.312 | 3.898 | 5.134 | 5.831 | 3.982 | 6.024 | 6.400 |
| n3n3' | 4.468 | 4.238 | 5.054 | 4.271 | 5.596 | 5.905 | 4.285 | 5.288 | 4.299 |

**Table 2.**

| 2-letter abbrev. | monomer screened for | # positive structures | percentage (out of 801) |
|---|---|---|---|
| A1 | complex A, monomer 1 | 9 | 1.1% |
| A2 | complex A, monomer 2 | 5 | 0.6% |
| B1 | complex B, monomer 1 | 12 | 1.5% |
| B2 | complex B, monomer 2 | 16 | 2.0% |
| C1 | complex C, monomer 1 | 1 | 0.1% |
| C2 | complex C, monomer 2 | 27 | 3.4% |
| D1 | complex D, monomer 1 | 8 | 1.0% |
| D2 | complex D, monomer 2 | 6 | 0.7% |
| H1 | complex H, monomer 1 | 8 | 1.0% |
| H2 | complex H, monomer 2 | 15 | 1.9% |
| I1 | complex I, monomer 1 | 0 | 0.0% |
| I2 | complex I, monomer 2 | 25 | 3.1% |
| P1 | complex P, monomer 1 | 8 | 1.0% |
| P2 | complex P, monomer 2 | 9 | 1.1% |
| Q1 | complex Q, monomer 1 | 5 | 0.6% |
| Q2 | complex Q, monomer 2 | 18 | 2.2% |
| Z1 | complex Z, monomer 1 | 6 | 0.7% |
| Z2 | complex Z, monomer 2 | 46 | 5.7% |

**Table 3.**

**A.**

**Complex A:**

| monomer 1 | | | | monomer 2 | | | |
|---|---|---|---|---|---|---|---|
| CPM | | TSM | | CPM | | TSM | |
| 1c1y:A | 1.8895 | 1c1y:A | 45.2035 | 1c1y:B | 9.5745 | 1c1y:B | 30.2432 |
| 1zsw:A | 4.6585 | 1m98:A | 13.5648 | 1o69:A | 11.7387 | 2a2o:F | 3.4961 |
| 1vp4:A | 11.4658 | 1m98:B | 14.4250 | 2f20:A | 16.0514 | 2f20:A | 17.7635 |
| 1sul:D | 11.4836 | 1vl4:B | 19.2273 | 2a2o:F | 23.1964 | 1o69:A | 36.2028 |
| 1sul:C | 11.6014 | 1nkv:C | 19.6930 | | | | |
| 1sul:A | 12.1042 | 1vgy:B | 20.5987 | | | | |
| 1sul:B | 12.1042 | 1sul:B | 21.6749 | | | | |
| 1vl4:B | 12.8482 | 1sul:D | 21.6833 | | | | |
| 1vgy:B | 13.6850 | 1sul:A | 21.7452 | | | | |
| 1nkv:C | 14.3992 | 1sul:C | 22.0641 | | | | |
| 1m98:B | 14.4657 | 1vp4:A | 23.3654 | | | | |
| 1m98:A | 14.5741 | 1zsw:A | 69.9124 | | | | |

**B.**

| | | | | | | |
|---|---|---|---|---|---|---|
| COMPLEX A, monomer 1: | 1m98 | D-6 | D-85 | S-10 | R-291 | AAAA |
| | 1m98 | D-85 | D-6 | S-10 | R-291 | BBBB |
| | 1nkv | D-94 | D-66 | S-47 | R-78 | CCCC |
| | 1su1 | D-9 | D-75 | S-76 | R-71 | AAAA |
| | 1su1 | D-75 | D-9 | S-76 | R-71 | BBBB |
| | 1su1 | D-75 | D-9 | S-76 | R-71 | CCCC |
| | 1su1 | D-75 | D-9 | S-76 | R-71 | DDDD |
| | 1vgy | D-138 | D-70 | S-18 | R-16 | BBBB |
| | 1vl4 | D-166 | D-146 | S-32 | R-34 | BBBB |
| | 1vp4 | D-111 | D-277 | S-275 | R-139 | AAAA |
| | 1zsw | D-187 | D-168 | S-172 | R-243 | AAAA |
| COMPLEX A, monomer 2: | 1o69 | T-328 | K-324 | R-329 | N-183 | AAAA |
| | 2a2o | T-13 | K-109 | R-12 | N-173 | FFFF |
| | 2f20 | T-80 | K-91 | R-77 | N-59 | AAAA |

**Table 4 (Part 1 of 2)**

## C. Determination of Inter-Monomer Overlap in Complex A ('Cutting Plane' Method)

| Monomer 1 | | | : | Monomer 2 | | | % Mon. 1 in Mon. 2 | % Mon. 2 in Mon. 1 |
|---|---|---|---|---|---|---|---|---|
| 1m98 | D-6 | AAAA | : | 1o69 | T-328 | AAAA | 30.36 | 9.64 |
| 1m98 | D-85 | BBBB | : | 1o69 | T-328 | AAAA | 29.87 | 9.64 |
| 1nkv | D-94 | CCCC | : | 1o69 | T-328 | AAAA | 5.88 | 9.64 |
| 1su1 | D-9 | AAAA | : | 1o69 | T-328 | AAAA | 24.29 | 9.64 |
| 1su1 | D-75 | BBBB | : | 1o69 | T-328 | AAAA | 24.36 | 9.64 |
| 1su1 | D-75 | CCCC | : | 1o69 | T-328 | AAAA | 24.29 | 9.64 |
| 1su1 | D-75 | DDDD | : | 1o69 | T-328 | AAAA | 24.36 | 9.64 |
| 1vgy | D-138 | BBBB | : | 1o69 | T-328 | AAAA | 20.79 | 9.64 |
| 1vl4 | D-166 | BBBB | : | 1o69 | T-328 | AAAA | 13.62 | 9.64 |
| 1vp4 | D-111 | AAAA | : | 1o69 | T-328 | AAAA | 32.58 | 9.64 |
| 1zsw | D-187 | AAAA | : | 1o69 | T-328 | AAAA | 2.45 | 9.64 |
| | | | | | | | | |
| 1m98 | D-6 | AAAA | : | 2a2o | T-13 | FFFF | 30.36 | 10.24 |
| 1m98 | D-85 | BBBB | : | 2a2o | T-13 | FFFF | 29.87 | 10.24 |
| 1nkv | D-94 | CCCC | : | 2a2o | T-13 | FFFF | 5.88 | 10.24 |
| 1su1 | D-9 | AAAA | : | 2a2o | T-13 | FFFF | 24.29 | 10.24 |
| 1su1 | D-75 | BBBB | : | 2a2o | T-13 | FFFF | 24.36 | 10.24 |
| 1su1 | D-75 | CCCC | : | 2a2o | T-13 | FFFF | 24.29 | 10.24 |
| 1su1 | D-75 | DDDD | : | 2a2o | T-13 | FFFF | 24.36 | 10.24 |
| 1vgy | D-138 | BBBB | : | 2a2o | T-13 | FFFF | 20.79 | 10.24 |
| 1vl4 | D-166 | BBBB | : | 2a2o | T-13 | FFFF | 13.62 | 10.24 |
| 1vp4 | D-111 | AAAA | : | 2a2o | T-13 | FFFF | 32.58 | 10.24 |
| 1zsw | D-187 | AAAA | : | 2a2o | T-13 | FFFF | 2.45 | 10.24 |
| | | | | | | | | |
| 1m98 | D-6 | AAAA | : | 2f20 | T-80 | AAAA | 30.36 | 26.62 |
| 1m98 | D-85 | BBBB | : | 2f20 | T-80 | AAAA | 29.87 | 26.62 |
| 1nkv | D-94 | CCCC | : | 2f20 | T-80 | AAAA | 5.88 | 26.62 |
| 1su1 | D-9 | AAAA | : | 2f20 | T-80 | AAAA | 24.29 | 26.62 |
| 1su1 | D-75 | BBBB | : | 2f20 | T-80 | AAAA | 24.36 | 26.62 |
| 1su1 | D-75 | CCCC | : | 2f20 | T-80 | AAAA | 24.29 | 26.62 |
| 1su1 | D-75 | DDDD | : | 2f20 | T-80 | AAAA | 24.36 | 26.62 |
| 1vgy | D-138 | BBBB | : | 2f20 | T-80 | AAAA | 20.79 | 26.62 |
| 1vl4 | D-166 | BBBB | : | 2f20 | T-80 | AAAA | 13.62 | 26.62 |
| 1vp4 | D-111 | AAAA | : | 2f20 | T-80 | AAAA | 32.58 | 26.62 |
| 1zsw | D-187 | AAAA | : | 2f20 | T-80 | AAAA | 2.45 | 26.62 |

**Table 4 (Part 2 of 2)**

COMPLEX B, monomer 1

| | | | | | |
|---|---|---|---|---|---|
| 1k77 | R-194 | S-146 | E-230 | Q-158 | AAAA |
| 1pf5 | R-26 | S-39 | E-3 | Q-36 | AAAA |
| 1s4k | R-11 | S-87 | E-38 | Q-104 | AAAA |
| 1s4k | R-11 | S-87 | E-38 | Q-104 | BBBB |
| 1tuh | R-68 | S-59 | E-49 | Q-74 | AAAA |
| 1tzz | R-1361 | S-1019 | E-1363 | Q-1351 | AAAA |
| 1wj9 | R-41 | S-208 | E-39 | Q-62 | AAAA |
| 1x72 | R-269 | S-25 | E-228 | Q-320 | AAAA |
| 2b3n | R-135 | S-90 | E-132 | Q-108 | AAAA |
| 2b3n | R-135 | S-90 | E-132 | Q-108 | BBBB |
| 2b4a | R-5 | S-93 | E-29 | Q-76 | AAAA |

COMPLEX B, monomer 2

| | | | | | |
|---|---|---|---|---|---|
| 1kuu | N-89 | E-36 | R-63 | R-91 | AAAA |
| 1o8c | N-93 | E-273 | R-55 | R-275 | BBBB |
| 1o8c | N-93 | E-273 | R-55 | R-275 | DDDD |
| 1rcu | N-148 | E-162 | R-150 | R-127 | AAAA |
| 1rcu | N-148 | E-162 | R-127 | R-150 | BBBB |
| 1rcu | N-148 | E-162 | R-127 | R-150 | CCCC |
| 1rcu | N-148 | E-162 | R-127 | R-150 | DDDD |
| 1rty | N-154 | E-28 | R-131 | R-151 | CCCC |
| 1rxd | N-23 | E-10 | R-46 | R-136 | AAAA |
| 1rxd | N-23 | E-10 | R-46 | R-136 | BBBB |
| 1rxd | N-23 | E-10 | R-46 | R-136 | CCCC |
| 1t0t | N-90 | E-31 | R-18 | R-33 | ZZZZ |
| 1tq8 | N-127 | E-88 | R-89 | R-147 | AAAA |
| 1vho | N-171 | E-108 | R-279 | R-100 | AAAA |
| 1vju | N-253 | E-48 | R-189 | R-269 | BBBB |
| 1vly | N-184 | E-92 | R-88 | R-9 | AAAA |
| 1yw1 | N-379 | E-283 | R-369 | R-3 | AAAA |
| 2ap6 | N-12 | E-91 | R-84 | R-9 | AAAA |
| 2ap6 | N-12 | E-91 | R-84 | R-9 | BBBB |
| 2ap6 | N-12 | E-91 | R-84 | R-9 | CCCC |
| 2ap6 | N-12 | E-91 | R-84 | R-9 | DDDD |
| 2ap6 | N-12 | E-91 | R-84 | R-9 | EEEE |
| 2ap6 | N-12 | E-91 | R-84 | R-9 | FFFF |
| 2ap6 | N-12 | E-91 | R-84 | R-9 | GGGG |
| 2ap6 | N-12 | E-91 | R-84 | R-9 | HHHH |
| 2b6c | N-137 | E-65 | R-138 | R-66 | AAAA |
| 2b6c | N-137 | E-65 | R-66 | R-138 | BBBB |

**Table 5 (Part 1 of 6)**

COMPLEX C, monomer 1

| | | | | | |
|---|---|---|---|---|---|
| 1xiz | R-74 | Y-72 | T-47 | H-5 | AAAA |
| 1xiz | R-74 | Y-72 | T-47 | H-5 | BBBB |

COMPLEX C, monomer 2

| | | | | | |
|---|---|---|---|---|---|
| 1ilv | P-122 | D-8 | D-9 | D-88 | BBBB |
| 1j9k | P-122 | D-88 | D-8 | D-9 | AAAA |
| 1j9l | P-122 | D-88 | D-8 | D-9 | AAAA |
| 1j9l | P-122 | D-8 | D-9 | D-88 | BBBB |
| 1l6r | P-185 | D-167 | D-8 | D-10 | AAAA |
| 1nf2 | P-179 | D-158 | D-8 | D-10 | AAAA |
| 1nf2 | P-779 | D-758 | D-814 | D-818 | CCCC |
| 1nkv | P-132 | D-105 | D-175 | D-176 | AAAA |
| 1nkv | P-132 | D-105 | D-175 | D-176 | BBBB |
| 1nkv | P-132 | D-105 | D-175 | D-176 | CCCC |
| 1pt8 | P-20 | D-128 | D-202 | D-43 | AAAA |
| 1rfz | P-82 | D-144 | D-80 | D-120 | AAAA |
| 1rfz | P-82 | D-80 | D-120 | D-144 | BBBB |
| 1rfz | P-82 | D-80 | D-120 | D-144 | DDDD |
| 1sd5 | P-115 | D-72 | D-75 | D-120 | AAAA |
| 1tel | P-1351 | D-1134 | D-1317 | D-1025 | AAAA |
| 1tel | P-2351 | D-2025 | D-2134 | D-2317 | BBBB |
| 1twu | P-55 | D-58 | D-122 | D-124 | AAAA |
| 1ufb | P-88 | D-60 | D-64 | D-93 | BBBB |
| 1ufb | P-88 | D-60 | D-64 | D-93 | DDDD |
| 1vgy | P-344 | D-134 | D-360 | D-70 | AAAA |
| 1vgy | P-344 | D-70 | D-134 | D-360 | BBBB |
| 1x7f | P-158 | D-160 | D-69 | D-99 | AAAA |
| 1xm7 | P-117 | D-159 | D-50 | D-80 | AAAA |
| 1xm7 | P-117 | D-50 | D-80 | D-159 | BBBB |
| 1xq6 | P-204 | D-179 | D-192 | D-193 | BBBB |
| 1y89 | P-124 | D-133 | D-154 | D-157 | BBBB |
| 1ydm | P-113 | D-141 | D-170 | D-117 | AAAA |
| 1ykw | P-351 | D-317 | D-25 | D-134 | AAAA |
| 1yw3 | P-165 | D-57 | D-60 | D-101 | BBBB |
| 1ze0 | P-66 | D-16 | D-18 | D-103 | AAAA |

**Table 5 (Part 2 of 6)**

COMPLEX D, monomer 1

| | | | | | |
|---|---|---|---|---|---|
| 1j2r | N-93 | Q-96 | E-75 | G-68 | AAAA |
| 1j2r | N-93 | Q-96 | E-75 | G-68 | BBBB |
| 1smb | N-91 | Q-94 | E-147 | G-143 | AAAA |
| 1wue | N-1302 | Q-1304 | E-1272 | G-1269 | AAAA |
| 1wue | N-2302 | Q-2304 | E-2272 | G-2269 | BBBB |
| 2f41 | N-268 | Q-265 | E-94 | G-92 | AAAA |
| 2f41 | N-268 | Q-265 | E-94 | G-92 | CCCC |

COMPLEX D, monomer 2

| | | | | | |
|---|---|---|---|---|---|
| 1sbk | R-27 | N-13 | S-44 | D-43 | AAAA |
| 1sbk | R-27 | N-13 | S-44 | D-43 | BBBB |
| 1sbk | R-27 | N-13 | S-44 | D-43 | CCCC |
| 1vh5 | R-27 | N-13 | S-44 | D-43 | AAAA |
| 1vi8 | R-27 | N-13 | S-44 | D-43 | AAAA |
| 1vi8 | R-27 | N-13 | S-44 | D-43 | EEEE |
| 1zbr | R-123 | N-139 | S-80 | D-118 | AAAA |
| 1zbr | R-123 | N-139 | S-80 | D-118 | BBBB |
| 2bdv | R-23 | N-21 | S-9 | D-13 | AAAA |

COMPLEX H, monomer 1

| | | | | | |
|---|---|---|---|---|---|
| 1ljo | K-16 | I-17 | E-34 | Q-73 | AAAA |
| 1vl4 | K-151 | I-152 | E-27 | Q-22 | AAAA |
| 1vl4 | K-151 | I-152 | E-27 | Q-22 | BBBB |
| 1woz | K-51 | I-42 | E-110 | Q-58 | AAAA |
| 1xvs | K-76 | I-75 | E-80 | Q-42 | AAAA |
| 1xvs | K-76 | I-75 | E-80 | Q-42 | BBBB |
| 1zsw | K-257 | I-251 | E-237 | Q-39 | AAAA |
| 1zxo | K-263 | I-264 | E-270 | Q-251 | FFFF |

COMPLEX H, monomer 2

| | | | | | |
|---|---|---|---|---|---|
| 1s7o | D-38 | E-43 | K-8 | D-46 | AAAA |
| 1sef | D-257 | E-204 | K-26 | D-27 | AAAA |
| 1v99 | D-60 | E-59 | K-102 | D-86 | CCCC |
| 1ve3 | D-93 | E-127 | K-102 | D-101 | AAAA |
| 1vju | D-25 | E-32 | K-230 | D-35 | AAAA |
| 1vju | D-25 | E-32 | K-230 | D-35 | BBBB |
| 1vjx | D-118 | E-57 | K-121 | D-60 | AAAA |
| 1vmh | D-124 | E-118 | K-73 | D-15 | AAAA |
| 1vpv | D-157 | E-282 | K-156 | D-258 | AAAA |
| 1wuf | D-1189 | E-1240 | K-1164 | D-1221 | AAAA |
| 1wuf | D-2189 | E-2240 | K-2164 | D-2221 | BBBB |
| 1xbf | D-15 | E-118 | K-73 | D-124 | BBBB |
| 1xg8 | D-49 | E-65 | K-48 | D-67 | AAAA |
| 1xmx | D-203 | E-205 | K-209 | D-208 | AAAA |
| 2b4w | D-53 | E-50 | K-133 | D-83 | AAAA |

**Table 5 (Part 3 of 6)**

COMPLEX I, monomer 1
<none detected>

COMPLEX I, monomer 2

| | | | | | |
|---|---|---|---|---|---|
| 1jog | Q-67 | N-9 | L-11 | L-139 | BBBB |
| 1nij | Q-274 | N-197 | L-195 | L-123 | AAAA |
| 1o65 | Q-122 | N-97 | L-155 | L-98 | AAAA |
| 1o65 | Q-122 | N-97 | L-98 | L-155 | BBBB |
| 1o65 | Q-122 | N-97 | L-98 | L-155 | CCCC |
| 1o67 | Q-122 | N-97 | L-155 | L-98 | AAAA |
| 1o67 | Q-122 | N-97 | L-98 | L-155 | BBBB |
| 1oq1 | Q-71 | N-131 | L-74 | L-114 | AAAA |
| 1oq1 | Q-71 | N-131 | L-74 | L-114 | DDDD |
| 1qy9 | Q-4 | N-29 | L-30 | L-50 | DDDD |
| 1qyi | Q-53 | N-77 | L-24 | L-81 | AAAA |
| 1sr0 | Q-172 | N-231 | L-177 | L-233 | AAAA |
| 1t06 | Q-110 | N-173 | L-136 | L-175 | AAAA |
| 1t06 | Q-110 | N-173 | L-136 | L-175 | BBBB |
| 1vgg | Q-126 | N-75 | L-5 | L-76 | AAAA |
| 1vgg | Q-126 | N-75 | L-5 | L-76 | BBBB |
| 1vgg | Q-126 | N-75 | L-5 | L-76 | CCCC |
| 1vgg | Q-126 | N-75 | L-5 | L-76 | DDDD |
| 1vkh | Q-218 | N-244 | L-150 | L-242 | AAAA |
| 1vkh | Q-218 | N-244 | L-150 | L-242 | BBBB |
| 1vp4 | Q-80 | N-244 | L-242 | L-100 | AAAA |
| 1vqr | Q-76 | N-200 | L-196 | L-57 | AAAA |
| 1vqr | Q-76 | N-200 | L-57 | L-196 | BBBB |
| 1vqr | Q-76 | N-200 | L-57 | L-196 | CCCC |
| 1vqr | Q-76 | N-200 | L-57 | L-196 | DDDD |
| 1x72 | Q-267 | N-315 | L-270 | L-317 | AAAA |
| 1y7m | Q-41 | N-30 | L-26 | L-33 | AAAA |
| 1ykw | Q-311 | N-111 | L-203 | L-262 | AAAA |
| 1ykw | Q-311 | N-111 | L-203 | L-262 | BBBB |
| 1yoz | Q-40 | N-15 | L-75 | L-83 | AAAA |
| 1yw3 | Q-102 | N-118 | L-108 | L-114 | DDDD |
| 1zee | Q-174 | N-306 | L-287 | L-307 | AAAA |
| 1zee | Q-174 | N-306 | L-287 | L-307 | BBBB |
| 1zxj | Q-60 | N-89 | L-20 | L-90 | DDDD |
| 1zzm | Q-171 | N-251 | L-197 | L-203 | AAAA |
| 2b6c | Q-182 | N-153 | L-122 | L-144 | AAAA |
| 2b6c | Q-182 | N-153 | L-122 | L-144 | BBBB |

**Table 5 (Part 4 of 6)**

COMPLEX P, monomer 1

| | | | | | |
|---|---|---|---|---|---|
| 1in0 | K-109 | S-78 | Q-154 | K-115 | AAAA |
| 1o8c | K-296 | S-178 | Q-316 | K-322 | BBBB |
| 1o8c | K-296 | S-178 | Q-316 | K-322 | DDDD |
| 1wdj | K-130 | S-121 | Q-122 | K-112 | AAAA |
| 1wuf | K-1220 | S-1022 | Q-1095 | K-1263 | AAAA |
| 1wuf | K-2220 | S-2022 | Q-2095 | K-2263 | BBBB |
| 2a8e | K-145 | S-95 | Q-174 | K-151 | AAAA |
| 2fds | K-195 | S-209 | Q-214 | K-277 | AAAA |
| 2fds | K-195 | S-209 | Q-214 | K-277 | BBBB |

COMPLEX P, monomer 2

| | | | | | |
|---|---|---|---|---|---|
| 1h2h | C-55 | S-156 | S-31 | S-57 | AAAA |
| 1pqy | C-24 | S-201 | S-20 | S-71 | AAAA |
| 1pt5 | C-22 | S-199 | S-18 | S-69 | AAAA |
| 1pt5 | C-22 | S-18 | S-69 | S-199 | BBBB |
| 1pt7 | C-22 | S-199 | S-18 | S-69 | AAAA |
| 1pt7 | C-22 | S-18 | S-69 | S-199 | BBBB |
| 1pt8 | C-22 | S-199 | S-18 | S-69 | AAAA |
| 1pt8 | C-22 | S-18 | S-69 | S-199 | BBBB |
| 1qy9 | C-72 | S-162 | S-53 | S-98 | AAAA |
| 1qy9 | C-72 | S-53 | S-98 | S-162 | DDDD |
| 1ri6 | C-248 | S-133 | S-200 | S-228 | AAAA |
| 1sdj | C-72 | S-162 | S-53 | S-98 | AAAA |

COMPLEX Q, monomer 1

| | | | | | |
|---|---|---|---|---|---|
| 1s4c | Y-102 | N-36 | K-61 | N-97 | AAAA |
| 1s4c | Y-102 | N-36 | K-61 | N-97 | BBBB |

COMPLEX Q, monomer 2

| | | | | | |
|---|---|---|---|---|---|
| 1j31 | S-98 | A-167 | E-42 | K-28 | CCCC |
| 1j31 | S-98 | A-167 | E-42 | K-28 | DDDD |
| 1oy1 | S-20 | A-140 | E-24 | K-6 | CCCC |
| 1ufa | S-403 | A-311 | E-184 | K-163 | AAAA |
| 1vgy | S-289 | A-60 | E-266 | K-279 | BBBB |
| 1ylo | S-10 | A-226 | E-21 | K-35 | AAAA |
| 1yzy | S-253 | A-149 | E-355 | K-340 | DDDD |
| 1z40 | S-161 | A-238 | E-159 | K-370 | AAAA |

**Table 5 (Part 5 of 6)**

COMPLEX Z, monomer 1

| | | | | | |
|---|---|---|---|---|---|
| 1in0 | N-159 | E-128 | R-131 | D-5 | AAAA |
| 1in0 | N-159 | E-128 | R-131 | D-5 | BBBB |
| 1s5a | N-35 | E-113 | R-118 | D-33 | CCCC |
| 1tuh | N-50 | E-125 | R-130 | D-48 | AAAA |
| 1xe7 | N-98 | E-50 | R-76 | D-175 | AAAA |
| 1xe7 | N-98 | E-50 | R-76 | D-175 | BBBB |
| 1xe7 | N-98 | E-50 | R-76 | D-175 | CCCC |
| 1xe8 | N-98 | E-50 | R-76 | D-175 | BBBB |
| 1xe8 | N-98 | E-50 | R-76 | D-175 | CCCC |

COMPLEX Z, monomer 2

| | | | | | |
|---|---|---|---|---|---|
| 1in0 | E-40 | K-44 | E-95 | K-49 | BBBB |
| 1ixl | E-33 | K-82 | E-110 | K-86 | AAAA |
| 1j31 | E-15 | K-25 | E-10 | K-12 | AAAA |
| 1j31 | E-10 | K-12 | E-15 | K-25 | BBBB |
| 1j31 | E-10 | K-12 | E-15 | K-25 | CCCC |
| 1j31 | E-10 | K-12 | E-15 | K-25 | DDDD |
| 1nf2 | E-686 | K-682 | E-738 | K-689 | CCCC |
| 1nig | E-125 | K-118 | E-129 | K-134 | AAAA |
| 1o61 | E-319 | K-322 | E-327 | K-382 | AAAA |
| 1o62 | E-319 | K-322 | E-327 | K-382 | AAAA |
| 1p1m | E-333 | K-315 | E-15 | K-10 | AAAA |
| 1rtw | E-9 | K-184 | E-189 | K-192 | BBBB |
| 1s12 | E-58 | K-42 | E-63 | K-55 | AAAA |
| 1s12 | E-158 | K-142 | E-163 | K-155 | BBBB |
| 1s12 | E-258 | K-242 | E-263 | K-255 | CCCC |
| 1s12 | E-358 | K-342 | E-363 | K-355 | DDDD |
| 1s4c | E-51 | K-60 | E-79 | K-145 | BBBB |
| 1t95 | E-62 | K-56 | E-23 | K-59 | AAAA |
| 1tpx | E-10 | K-115 | E-16 | K-12 | BBBB |
| 1tqb | E-10 | K-115 | E-16 | K-12 | BBBB |
| 1tqc | E-10 | K-115 | E-16 | K-12 | BBBB |
| 1ue8 | E-213 | K-141 | E-116 | K-348 | AAAA |

COMPLEX Z, monomer 2 (cont'd.)

| | | | | | |
|---|---|---|---|---|---|
| 1vl4 | E-162 | K-150 | E-27 | K-7 | AAAA |
| 1vl4 | E-27 | K-7 | E-162 | K-150 | BBBB |
| 1vl5 | E-202 | K-215 | E-213 | K-223 | BBBB |
| 1vl5 | E-202 | K-215 | E-213 | K-223 | CCCC |
| 1vl5 | E-202 | K-215 | E-213 | K-223 | DDDD |
| 1vph | E-69 | K-66 | E-72 | K-78 | DDDD |
| 1vpv | E-38 | K-52 | E-43 | K-266 | BBBB |
| 1vqw | E-368 | K-371 | E-375 | K-401 | AAAA |
| 1vqw | E-368 | K-371 | E-375 | K-401 | BBBB |
| 1wdt | E-319 | K-287 | E-366 | K-367 | AAAA |
| 1wk4 | E-132 | K-131 | E-150 | K-159 | BBBB |
| 1wr2 | E-54 | K-53 | E-84 | K-82 | AAAA |
| 1wu8 | E-51 | K-99 | E-189 | K-174 | BBBB |
| 1wu8 | E-51 | K-99 | E-189 | K-174 | CCCC |
| 1wvi | E-1183 | K-1189 | E-1183 | K-1217 | AAAA |
| 1wvi | E-2183 | K-2189 | E-2183 | K-2217 | BBBB |
| 1wvi | E-3183 | K-3189 | E-3183 | K-3217 | CCCC |
| 1wvi | E-4183 | K-4189 | E-4183 | K-4217 | DDDD |
| 1xg7 | E-14 | K-17 | E-20 | K-232 | BBBB |
| 1xx7 | E-104 | K-59 | E-116 | K-121 | DDDD |
| 1xx7 | E-104 | K-59 | E-116 | K-121 | FFFF |
| 1ydw | E-86 | K-90 | E-93 | K-113 | BBBB |
| 1yf9 | E-43 | K-45 | E-53 | K-159 | BBBB |
| 1yf9 | E-43 | K-45 | E-53 | K-159 | CCCC |
| 1ytl | E-161 | K-56 | E-169 | K-165 | BBBB |
| 1yzy | E-267 | K-260 | E-405 | K-268 | AAAA |
| 1zxj | E-103 | K-31 | E-136 | K-140 | BBBB |
| 2a8e | E-179 | K-177 | E-8 | K-33 | AAAA |
| 2aca | E-138 | K-20 | E-171 | K-16 | AAAA |
| 2aca | E-138 | K-16 | E-171 | K-20 | BBBB |
| 2ap3 | E-32 | K-90 | E-124 | K-125 | AAAA |
| 2auw | E-28 | K-82 | E-87 | K-126 | BBBB |
| 2b8m | E-56 | K-15 | E-94 | K-54 | AAAA |
| 2f4n | E-212 | K-172 | E-262 | K-204 | CCCC |

**Table 5 (Part 6 of 6)**

| PDB ID | Interface 3D SM Detected | PDB Header | Source Organism | Remarks | Published Reference |
|---|---|---|---|---|---|
| 1ZSW[a] | complex A[c], monomer 1 | STRUCTURAL GENOMICS, UNKNOWN FUNCTION 25-MAY-05 | *Bacillus cereus* | METALLO PROTEIN FROM GLYOXALASE FAMILY | none |
| 1NKV | complex A, monomer 1 | STRUCTURAL GENOMICS, UNKNOWN FUNCTION 03-JAN-03 | *Escherichia coli* | E COLI YJHP GENE PRODUCT NESG TARGET ER13 | none |
| 1O69 | complex A, monomer 2 | STRUCTURAL GENOMICS, UNKNOWN FUNCTION 23-OCT-03 | *Campylobacter jejuni* | PRESUMPTIVE PLP-DEPENDENT AMINOTRANSFERASE | Proteins 2005 vol. 60, pp. 787ff |
| 2A2O | complex A, monomer 2 | HYPOTHETICAL PROTEIN, TRANSCRIPTION 22-JUN-05 | *Bacteroides thetaiotaomicron VPI-5482* | PUTATIVE TEN-A FAMILY TRANSCRIPTIONAL REGULATOR | none |
| 1K77 | complex B[d], monomer 1 | STRUCTURAL GENOMICS, UNKNOWN FUNCTION 18-OCT-01 | *Escherichia coli* | GLYOXYLATE-INDUCED HYPOTHETICAL PROTEIN YGBM (EC1530) | Proteins. 2002 Aug 1;48(2):427-30 |
| 1KUU | complex B, monomer 2 | STRUCTURAL GENOMICS, UNKNOWN FUNCTION 22-JAN-02 | *Methanothermobacter thermautotrophicus* | CONSERVED PROTEIN MTH1020 WITH NTN-HYDROLASE FOLD | none |
| 2AP6 | complex B, monomer 2 | UNKNOWN FUNCTION 15-AUG-05 | *Agrobacterium tumefaciens* | HYPOTHETICAL PROTEIN ATU4242, NESG TARGET ATR43 | none |
| 1XIZ | complex C[e], monomer 1 | STRUCTURAL GENOMICS, UNKNOWN FUNCTION 22-SEP-04 | *Salmonella typhimurium* | PUTATIVE MANNITOL/FRUCTOSE-SPECIFIC PHOSPHOTRANSFERASE SYSTEM, DOMAIN IIA | none |
| 1NKV | complex C, monomer 2 | STRUCTURAL GENOMICS, UNKNOWN FUNCTION 03-JAN-03 | *Escherichia coli* | HYPOTHETICAL PROTEIN YJHP, NESG TARGET ER13 | none |
| 1J2R | complex D[f], monomer 1 | STRUCTURAL GENOMICS, UNKNOWN FUNCTION 09-JAN-03 | *Escherichia coli* | HYPOTHETICAL ISOCHORISMATASE FAMILY PROTEIN YECD WITH PARALLEL BETA-SHEET 3-2-1-4-5-6, ALPHA-BETA-ALPHA MOTIF | Nucleic Acids Res. 2004 Jul 1;32(Web Server issue):W606-9 |
| 2F4L | complex D, monomer 1 | STRUCTURAL GENOMICS, UNKNOWN FUNCTION 23-NOV-05 | *Thermotoga maritima* | PROTEIN TM0119, A PUTATIVE ACETAMIDASE | none |
| 1SBK | complex D, monomer 2 | STRUCTURAL GENOMICS, UNKNOWN FUNCTION 10-FEB-04 | *Escherichia coli* | HYPOTHETICAL PROTEIN YDII, NESGC TARGET ER29 | none |
| 1VH5 | complex D, monomer 2 | STRUCTURAL GENOMICS, UNKNOWN FUNCTION 01-DEC-03 | *Escherichia coli* | HYPOTHETICAL PROTEIN YDII, A PUTATIVE THIOESTERASE | none |
| 1VI8 | complex D, monomer 2 | STRUCTURAL GENOMICS, UNKNOWN FUNCTION 01-DEC-03 | *Escherichia coli* | HYPOTHETICAL PROTEIN YDII, A PUTATIVE THIOESTERASE | none |
| 1ZBR | complex D, monomer 2 | STRUCTURAL GENOMICS, UNKNOWN FUNCTION 08-APR-05 | *Porphyromonas gingivalis* | A CONSERVED HYPOTHETICAL ALPHA-BETA PROTEIN, PUTATIVE ARGININE DEIMINASE | none |
| 1XVS | complex H[g], monomer 1 | STRUCTURAL GENOMICS, UNKNOWN FUNCTION 28-OCT-04 | *Vibrio cholerae* | APAG PROTEIN | none |
| 1S7O | complex H, monomer 2 | STRUCTURAL GENOMICS, UNKNOWN FUNCTION 29-JAN-04 | *Streptococcus pygenes* | HYPOTHETICAL PROTEIN UPF0122, PUTATIVE DNA BINDING PROTEIN SP_1288; SRP, RNA POLYMERASE SIGMA FACTOR | Acta Crystallogr D Biol Crystallogr. 2004 Jul;60(Pt 7):1266-71. Epub 2004 Jun 22 |
| 1V99 | complex H, monomer 2 | STRUCTURAL GENOMICS, UNKNOWN FUNCTION 23-JAN-04 | *Pyrococcus horikoshii OT3* | PERIPLASMIC DIVALENT CATION TOLERANCE PROTEIN CUTA, WITH CUCL2 | none |
| 1XG8 | complex H, monomer 2 | STRUCTURAL GENOMICS, UNKNOWN FUNCTION 16-SEP-04 | *Staphylococcus aureus (APC23712)* | HYPOTHETICAL PROTEIN SA0789 | none |
| 1IN0 | complex P[h], monomer 1 | STRUCTURAL GENOMICS, UNKNOWN FUNCTION 11-MAY-01 | *Haemophilus influenzae* | YAJQ PROTEIN, W/ ALPHA AND BETA SANDWICH MOTIF & TANDEM OF 3 RNP-LIKE DOMAINS | none |
| 1SDJ | complex P, monomer 2 | STRUCTURAL GENOMICS, UNKNOWN FUNCTION 13-FEB-04 | *Pseudomonas fluorescens* | HYPOTHETICAL PROTEIN YDDE (SYN.: ORFB, NESGC TARGET ET25), PHENAZINE BIOSYNTHETIC PROTEIN PHZF | none |
| 1S4C | complex Q[i], monomer 1 | STRUCTURAL GENOMICS, UNKNOWN FUNCTION 15-JAN-04 | *Haemophilus influenzae* | YHCH PROTEIN (HI0227) COPPER COMPLEX W/ DOUBLE-STRANDED BETA-HELIX | J Bacteriol. 2005 Aug;187(16):5520-7 |
| 1UFA | complex Q, monomer 2 | STRUCTURAL GENOMICS, UNKNOWN FUNCTION 28-MAY-03 | *Thermus thermophilus HB8* | PROTEIN TT1467 | none |
| 1TUH | complex Z[j], monomer 1 | UNKNOWN FUNCTION 25-JUN-04 | UNCULTURED BACTERIUM | HYPOTHETICAL PROTEIN EGC068, NEW MEMBER OF THE ADAPTABLE ALPHA+BETA BARREL FAMILY (SYN.: BAL32A) | J Mol Biol. 2005 Mar 11;346(5):1229-41. Epub 2004 Dec 31 |

**Table 6. (part 1 of 2)**

| | | | | | |
|---|---|---|---|---|---|
| 1NIG | complex Z, monomer 2 | STRUCTURAL GENOMICS, UNKNOWN FUNCTION 23-DEC-02 | *Thermoplasma acidophilum* | HYPOTHETICAL PROTEIN TA1238 W/ 4 HELIXBUNDLES, A NEW TYPE OF HELICAL SUPER-BUNDLE | none |
| 1P1M | complex Z, monomer 2 | STRUCTURAL GENOMICS, UNKNOWN FUNCTION 12-APR-03 | *Thermotoga maritima* | HYPOTHETICAL PROTEIN TM0936, W/ BOUND NI AND METHIONINE | none |
| 1VPV | complex Z, monomer 2 | STRUCTURAL GENOMICS, UNKNOWN FUNCTION 18-NOV-04 | *Thermotoga maritima* | UPF0230 PROTEIN TM1468 | none |
| 1WDT[b] | complex Z, monomer 2 | STRUCTURAL GENOMICS, UNKNOWN FUNCTION 17-MAY-04 | *Thermus thermophilus HB8* | PROTEIN TTK003000868, ELONGATION FACTOR G HOMOLOG, GTP COMPLEX | none |

[a] A heme-NO (nitric oxide) 3D SM has also been detected in this structure; see Reyes, V.M. (2008b).

[b] A heme-NO (nitric oxide) 3D SM has also been detected in this structure; see Reyes, V.M. (2008b).

[c] Complex A is RAP•Gmppnp complexed w/ c-RAF1 Ras-binding domain (RAFRBD); RAP is monomer 1; RAFRBD is monomer 2

[d] Complex B is RHOA complexed w/ protein kinase PKN/PRK1 effector domain (PK PKN/PRK1 ED); RHOA is monomer 1; PK PKN/PRK1 ED is monomer 2

[e] Complex C is RAC complexed w/ RHOGD1; RAC is monomer 1; RHOGD1 is monomer 2

[f] Complex D is RAC complexed w/ P67PHOX; RAC is monomer 1; P67PHOX is monomer 2

[g] Complex H is kinase-associated phosphatase (KAP) complexed w/ phospho-CDK2; KAP is monomer 1; phospho-CDK2 is monomer 2

[h] Complex P is an immunoglobulin light chain dimer (Bence-Jones protein); monomers 1 and 2 are both IgG light chains

[i] Complex Q is beta-catenin complexed with HTCF-4; beta-catenin is monomer 1; HTCF-4 is monomer 2

[j] Complex Z is an interleukin-2 homodimer; monomers 1 and 2 are both IL-2s

**Table 6. (part 2 of 2)**